\documentclass[aps, prb, superscriptaddress, reprint, onecolumn]{revtex4-2}

\usepackage{amsmath}
\usepackage{amsfonts}
\usepackage{amssymb}
\usepackage{gensymb}
\usepackage{graphicx}
\usepackage[utf8]{inputenc}
\usepackage[colorlinks=true, linkcolor=black, urlcolor=blue, citecolor=black, anchorcolor=black]{hyperref}
\usepackage{enumitem}

\usepackage{fancyvrb}
\usepackage{algorithm}

\setlist{nosep}

\DeclareMathOperator\erf{erf}
\DeclareMathOperator\erfc{erfc}

\begin{document}
\title{Rapid and facile reconstruction of time-resolved fluorescence data with exponentially modified Gaussians}
\author{Darien J. Morrow}
\email{darienmorrow@gmail.com}
\affiliation{Center for Nanoscale Materials, Argonne National Laboratory, Lemont, Illinois 60439, United States}

\author{Xuedan Ma}
\email{xuedan.ma@anl.gov}
\affiliation{Center for Nanoscale Materials, Argonne National Laboratory, Lemont, Illinois 60439, United States}
\affiliation
{Consortium for Advanced Science and Engineering, University of Chicago, Chicago, Illinois 60637, United States}
\affiliation
{Northwestern-Argonne Institute of Science and Engineering, 2205 Tech Drive, Evanston, IL 60208, USA}

\date{\today}

\begin{abstract}
	Analyte response is convoluted with instrument response in time resolved fluorescence data. 
	Decoding the desired analyte information from the measurement usually requires  iterative numerical convolutions.
	Here in, we show that time resolved data can be completely, analytically reconstructed without numerical convolutions.
	Our strategy relies on a summation of exponentially modified Gaussians which encode all convolutions within easily evaluated complementary error functions.  
	Compared to a numerical convolution strategy implemented with Python, this new method is computationally cheaper and scales less steeply with the number of temporal points in the experimental dataset.
\end{abstract}

\maketitle

\section{Introduction}

\begin{figure}[!htbp]
	\centering
	\includegraphics[trim=0cm 1cm 2cm 2cm, clip,width=0.5\linewidth]{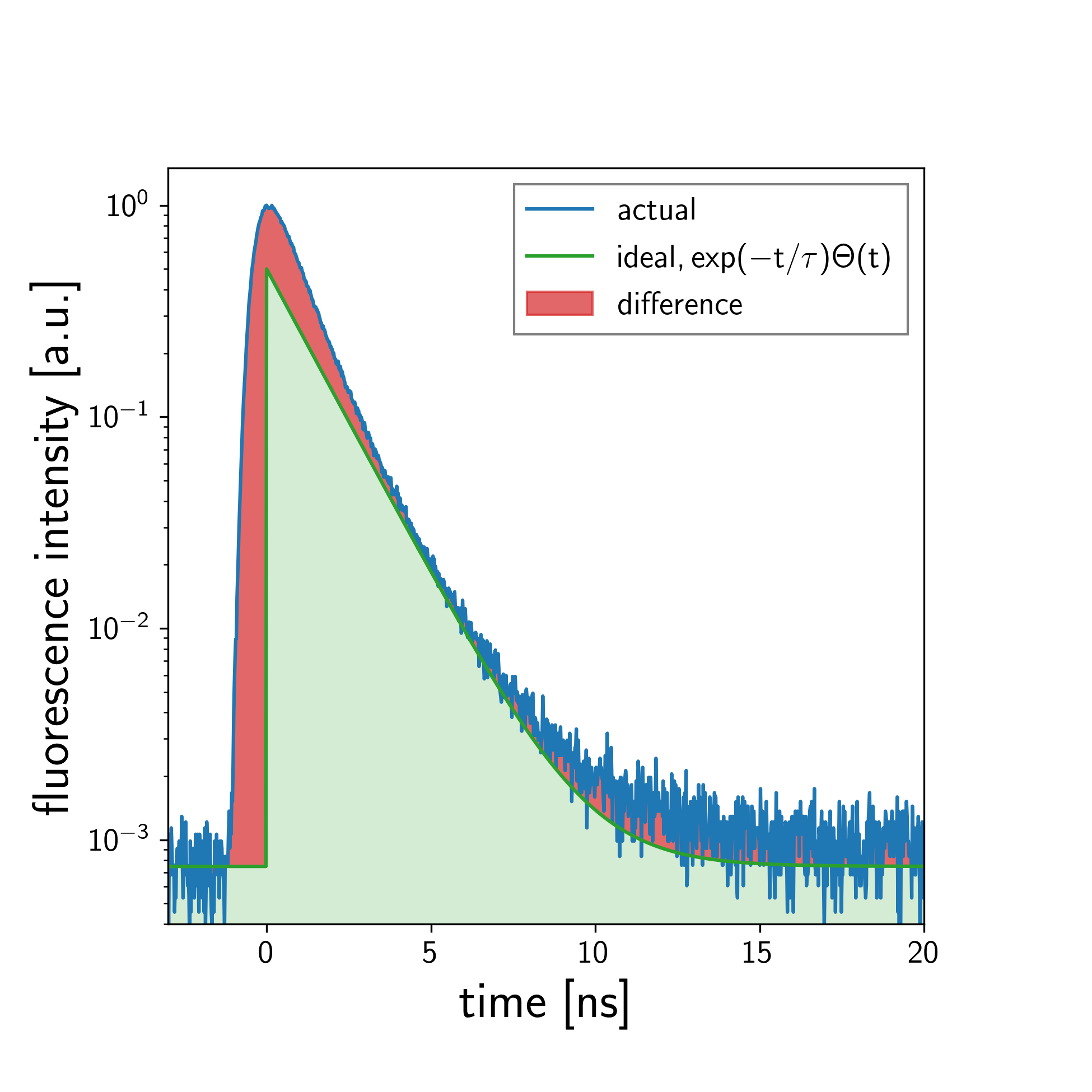}
	\caption{
		Actual time-resolved flourescence data, $M$, versus the ideal exponential decay predicted from first order kinetics, $S$. $\tau =1/\lambda = 1.5 \text{ ns}$.
	}\label{fig0:ideality} 
\end{figure}

Time resolved fluorescence is a ubiquitous measurement used in fields as diverse as materials physics,\cite{Fu_Jin_2017} physical chemistry,\cite{Goldsmith_Moerner_2010} structural biology,\cite{Elson_French_2004, Bastiaens_Squire_1999} robotic surgery,\cite{Gorpas_Farwell_2019} and art conservation.\cite{Comelli_Toniolo_2004} 
In these uses, the metric of interest is the set of fluorescence lifetimes, $\{\tau_i\}$, which is defined by the physical identity and structure of the sample. 
In the Platonic ideal of the experiment, a sample is excited with an infinitely fast pulse of light, the sample then undergoes spontaneous emission with first order kinetics, so the emitted intensity of light decays exponentially with time.
The emitted light is finally measured with an infinitely fast detector. 
In actuality, excitation pulses have finite width and detectors have finite response times (\autoref{fig0:ideality}). 
These non-idealities must be accounted for in data analysis routines to extract useful and representative fluorescence lifetimes. 

If $S$ is the ideal signal, the measured response, $M$, is the convolution of the ideal signal with some instrument response function, $R$, which accounts for the non-idealities of reality,
\begin{align}
M&=S*R.
\end{align}
Deconvolution of $S$ from $M$ can be accomplished if $R$ is exactly known. 
If $R$ is not known with certainty, then deconvolution is a non-covex (usually unfeasible) optimization problem.\cite{Knight_Selinger_1971} 
Instead most commercial software packages and individual researchers rely on a reconvolution strategy in which a decay model, $S_{\text{model}}$, is convolved with a measured (or supposed) instrument response function, $R_{\text{measured}}$,
\begin{align}
	M_{\text{model}}&=S_{\text{model}}*R_{\text{measured}}.
\end{align}
The resultant signal, $M_{\text{model}}$, is compared to $M$. 
The parameters which define $S_{\text{model}}$ are then iteratively adjusted to minimize the difference between $M_{\text{model}}$ and $M$ in order to extract a representation of $S$. 
In the case of methods like Fluorescence Lifetime Imaging (FLIM), the extraction of $S$ must then be accomplished for thousands of time traces which each represents a pixel of a spatial map.\cite{Becker_2012}

The reconvolution strategy is usually robust and generates an accurate view of the dynamics encoded in fluorescence data. 
In practice, it can be difficult to correctly zero-pad and interpolate $S_{\text{model}}$ and $R$. 
Moreover, if one does not use an easily scriptable analysis software, then one must iterate the parameters defining $S_{\text{model}}$ manually, which can be time consuming. 
Closed source commercial software packages (e.g. \texttt{SPCImage NG} from Becker \& Hickl and \texttt{EasyTau 2} from Picoquant) and open source packages (e.g. \texttt{DecayFit} and \texttt{FLIMfit},\cite{Warren_Degtyar_2013}) exist to help solve the problems associated with fitting fluorescence data to extract lifetimes via numerical convolutions.
Without using these packages, in many cases, researchers choose to assume the mid to late time dynamics encoded in their datasets are representative of $S$ and therefore just fit $S_{\text{model}}$ to $M\left(t>t_{\text{cutoff}}\right)$, which is a good assumption in some limits. 
It is likely that this choice is oftentimes made not for suitability of assumption, but instead because it can be hard to correctly implement numerical convolutions.  

Herein we introduce an analytic solution to this numerical convolution problem. 
We derive an integral equation which accounts for any number of exponential decay processes in both the sample response and instrument response. 
All numerically difficult convolutions are accounted for analytically. 
Our result is constructed so that analysis of time-resolved fluorescence data exclusively uses the robust algorithms which are packaged in most software packages to calculate the values of exponential and error functions. 
We demonstrate the utility of our result for reconstructing the wavelength dependent response of a commercial single-photon avalanche diode (SPAD) and also for retrieving the decay lifetimes of a model system.
Finally we show that our method is computationally cheaper than more established methods.

\section{Derivation of primary result}

In this section we derive our primary result (\autoref{eq:bigresult}). 

\subsection{Exponentially modified Gaussian: h(x)}

We start by assuming that fluorescence decay processes can be described using monoexponential decays 
\begin{align}
	f(x; \lambda) &\equiv \lambda \exp{\left[-\lambda x\right]}\Theta\left[x\right], \label{eq:f} \\
	\Theta\left[x\right] &\equiv
	\begin{cases}
		1, x>0 \\
		0, x\leq0
	\end{cases}.
\end{align}
in which $\Theta$ is the Heaviside step function, which accounts for the sample not emitting until after it is excited.
The lifetime of a component is given by $\tau_i = \frac{1}{\lambda_i}$.
The ideal instrument response function (IRF) is considered to be a Gaussian
\begin{align}
	g(x; \mu, \sigma) &\equiv \frac{1}{\sigma \sqrt{2\pi}} \exp{\left[-\left(\frac{x-\mu}{\sqrt{2}\sigma}\right)^2\right]} \label{eq:g},
\end{align}
with $\sigma$ characterizing the temporal width and $\mu$ defining the ``time zero'' of the instrument.
Later we will account for non-Gaussian IRFs.
Note that, for clarity, $f(x)$ and $g(x)$ are both written as normalized functions so that  $\int_{-\infty}^{\infty}f(x)\textrm{d}x =1$---this choice does not affect the form of our final result. 

We now define a composite function which will be the mainstay of our derivation,
\begin{align}
	h(x; \mu, \sigma, \lambda) &\equiv f(x; \lambda) * g(x; \mu, \sigma) \label{eq:h},\\
	(f * g)(t) &\equiv \int_{-\infty}^\infty f(\tau) g(t - \tau) \textrm{d}\tau. 
\end{align}
Here, $*$ is the convolution operator which, importantly, is commutative, associative,  and distributive.
In Appendix \ref{Ap2} we show that $h$ is a function known as the \emph{exponentially modified Gaussian} (EMG) with form,
\begin{align}
	\begin{split}
	h(x; \mu, \sigma, \lambda) =  \frac{\lambda}{2}\exp{\left[\frac{\lambda}{2}\left(2\mu + \lambda \sigma^2 - 2x\right)  \right]} \\ \times \erfc{\left[\frac{\mu + \lambda\sigma^2 - x}{\sqrt{2}\sigma}\right]}, \label{eq:emg}
	\end{split}
\end{align}
where
\begin{align}
	\erfc{(t)} \equiv 1 - \erf{(t)} = \frac{2}{\sqrt{\pi}}\int_{t}^{\infty}\exp{\left[-\tau^2\right]}\textrm{d}\tau. \label{eq:erfc}
\end{align}
In \autoref{eq:emg}, the convolution operation present in \autoref{eq:h} is rewritten as a single \emph{complementary error} function.
The EMG is commonly used in many fields as a phenomenological quantification tool: 
in the chromatography and mass spectrometry fields it is used to quantify tailed lineshapes,\cite{Jeansonne_Foley_1991, Kalambet_Tikhonov_2011, Purushothaman_Yavor_2017} in psychophysiology it is used to quantify response times,\cite{Golubev2017, Matzke2009} and in material science it has seen limited use fitting fluorescence spectra and diffusion profiles from inhomogeneous semiconductors.\cite{Pan_Jin_2020, Elbaz2017, Lockwood2004, Ardekani2019, Yoo2015}

\begin{figure}[!htbp]
	\centering
	\includegraphics[trim=0cm 1cm 2cm 2cm, clip,width=0.5\linewidth]{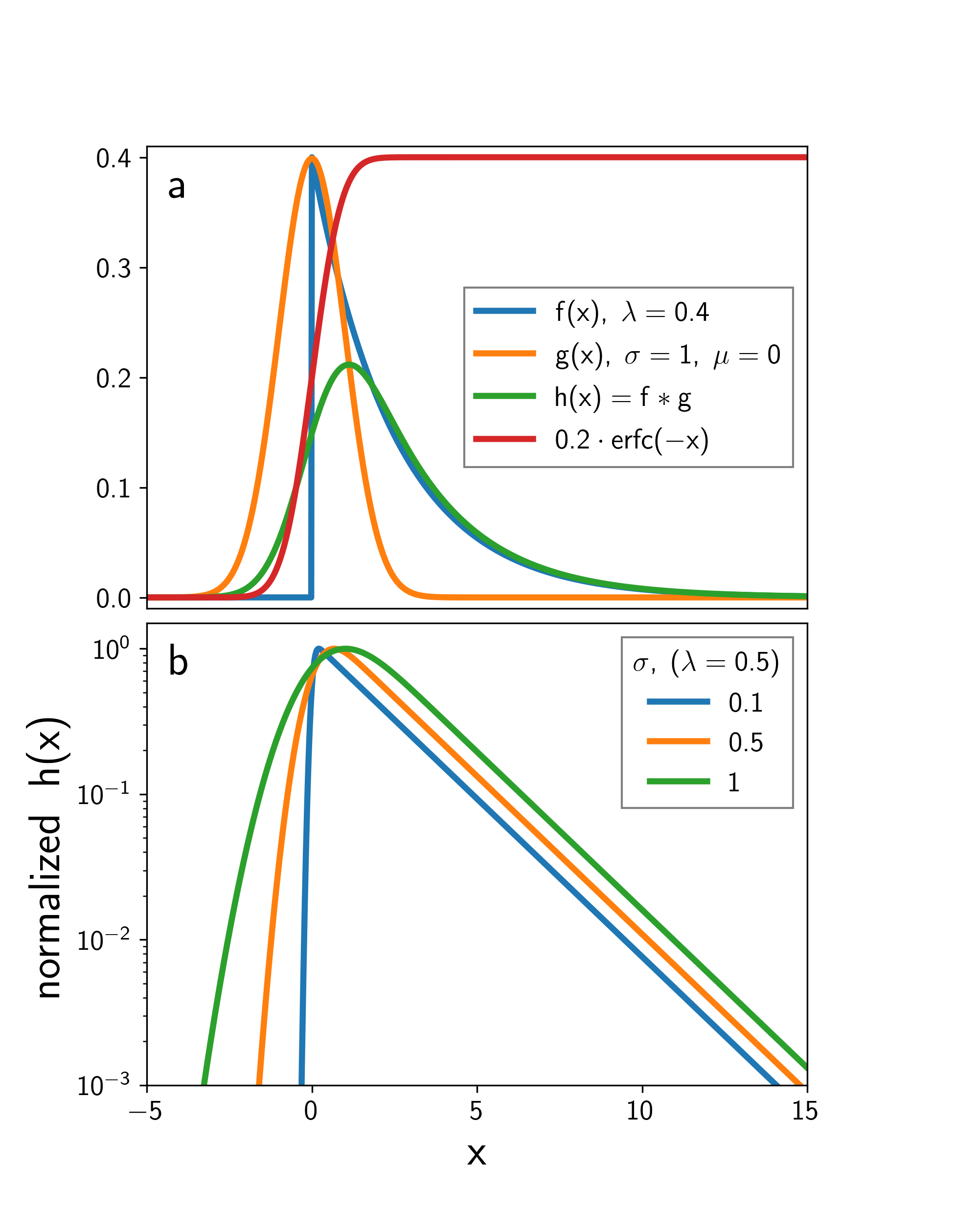}
	\caption{Graphical form of the exponentially modified Gaussian. (a)	Graph of \autoref{eq:f}, \autoref{eq:g}, and \autoref{eq:h} for $\lambda=0.4$, $\sigma=1$ and $\mu=0$. A scaled form of \autoref{eq:erfc} is also shown to highlight its sigmoid-like behavior. (b) \autoref{eq:h} with $\lambda=0.5$ and $\sigma \in \{0.1, 0.5, 1\}$ showing how larger values of $\sigma$ decreases the angle of $h$ at negative $x$ and pushes the maximum to positive $x$.  
	}\label{fig1:EMG} 
\end{figure}

\autoref{fig1:EMG}a graphs $f$, $g$, and $h$.
At early times $h$ is similar to the Gaussian IRF, $g$, but at late times, $h$ follows the exponential decay, $f$. 
\autoref{fig1:EMG} validates the intuition that if the dynamics of interest are much longer than the width of the IRF, then one can extract the lifetimes of interest by fitting the long-time tail of the measured response. 
\autoref{fig1:EMG}b shows how larger widths of the IRF ($\sigma$) pushes the peak maximum of $h$ to later times. 

\subsection{Extension of h(x) with multiple decay functions}

For multiple decay pathways we write a distribution of exponential decays, 
\begin{align}
	F^{(i)}(x) &\equiv \sum_{i}^{m} a_i \lambda_i \exp{\left[-\lambda_i x\right]}\Theta\left[x\right], \label{eq:F}\\
&= \sum_{i}^{m} a_i f_i(x),\label{eq:exps}
\end{align}
with $a_i$ being a weighting factor for each exponential decay with $\alpha_i\equiv a_i\lambda_i$ being the unnormalized weight.  
Convolving this distribution of decays with a Gaussian yields
\begin{align}
	H^{(i)}(x) &\equiv g(x) * F^{(i)}(x), \\
	& = \sum_{i}^{m}  g(x) * \left[a_i f_i(x)\right], \\
	& = \sum_{i}^{m}  a_i h_i(x), \label{eq:H}
\end{align}
in which we noted that convolution is a linear operation and therefore distributive.
Here we emphasize that $h_i(x) = h(x; \mu, \sigma, \lambda_i)$ which means that there are single $\mu$ and $\sigma$ but multiple $\lambda_i$. 
If a time-resolved fluorescence experiment has a Gaussian IRF, then $H^{(i)}(x)$ describes the measured response of a sample with a set, $\{\lambda_i\}$, of decay pathways.

\subsection{Detectors with non-Gaussian, tailed IRFs}

Oftentimes SPAD detectors do \emph{not} have Gaussian IRFs, but instead have a long tail towards positive times. Phenomenologically these non-ideal IRFs may be described with a distribution of weights and rates.
In this case we interpret $H^{(i)}(x)$ as the IRF.
We then account for analyte response by convolving $H^{(i)}(x)$ (our IRF) with a second $F^{(j)}(x)$ (the true sample decay pathways) to form a representative measured signal, $\mathcal{M}$,
\begin{align}
	\mathcal{M}(x) &= H^{(i)}(x) * F^{(j)}(x), \label{eq:Msmall}
\end{align}
which may be rewritten as 
\begin{align}
	\mathcal{M}(x) & = g(x) * F^{(i)}(x) * F^{(j)}(x), \\
	&= g(x) * \left\{\sum_{i}^{m}  f_i(x)\right\} * \left\{\sum_{j}^{n}  f_j(x)\right\}, \\
	&= g(x) * \left\{\sum_{i,j}^{m,n}  f_i(x) *  f_j(x)\right\}, \\
	&= g(x) * \left\{\sum_{i,j}^{m,n}  y_{i,j}(x)\right\}. \label{eq:M}
\end{align}
To solve the double convolution present in \autoref{eq:M} we first consider the argument of the summand inside the braces
\begin{align}
	y_{i,j}(t) &= f_i(t) * f_j(t), \\
	&= a_i \lambda_i a_j \lambda_j \int_{-\infty}^\infty e^{\left[-\lambda_i \tau\right]}\Theta\left[\tau\right] e^{\left[-\lambda_j (t - \tau)\right]}\Theta\left[t-\tau\right]\textrm{d}\tau.
\end{align}
This integrand is nonzero only when $\tau > 0$ and $t-\tau >0$ so 
\begin{align}
	y_{i,j}(t) &= a_i \lambda_i a_j \lambda_j \exp{\left[-\lambda_j \tau\right]} \Theta\left[t\right] \int_{0}^t \exp{\left[\tau(\lambda_j - \lambda_i)\right]}\textrm{d}\tau, \\
	&= \left(\frac{a_j\lambda_j}{\lambda_j - \lambda_i}\right) f_i(t) + \left(\frac{a_i\lambda_i}{\lambda_i - \lambda_j}\right) f_j(t), \label{eq:y}
\end{align}
in which we noted the standard integral 
\begin{align}
	\int_{0}^t \exp{\left[\tau(\lambda_j - \lambda_i)\right]}\textrm{d}\tau = \frac{1-\exp{\left[(\lambda_j-\lambda_i)t\right]}}{\lambda_j - \lambda_i}\text{, for } \lambda_j \neq \lambda_i.
\end{align}
\autoref{eq:y} is only valid when the sample's decay rates are not the same as the instrument response decay rates. Practically, this requirement is satisfied when the sample has slower dynamics than the detector.

Substitution of $y_{i,j}$ from \autoref{eq:y} into \autoref{eq:M} yields our measured response 
\begin{align}
	\mathcal{M}(x) &=  g(x) * \sum_{i,j}^{m,n} \left\{ \left(\frac{a_j\lambda_jf_i(x)}{\lambda_j - \lambda_i}\right)  + \left(\frac{a_i\lambda_if_j(x)}{\lambda_i - \lambda_j}\right) \right\}, \\
	&=  \sum_{i,j}^{m,n} \left\{ \left(\frac{a_j\lambda_j}{\lambda_j - \lambda_i}\right) h_i(x) + \left(\frac{a_i\lambda_i}{\lambda_i - \lambda_j}\right) h_j(x)\right\}, \\
	&= \sum_i^m h_i(x) \sum_{j}^{n}\frac{a_j\lambda_j}{\lambda_j - \lambda_i} + \sum_j^n h_j(x) \sum_{i}^{m}\frac{a_i\lambda_i}{\lambda_i - \lambda_j} \label{eq:bigresult}.
\end{align}
\autoref{eq:bigresult} is our primary analytic result. 
It encodes the dynamics of an experiment whose IRF has an arbitrarily large number of exponential tails, and whose sample response also has an arbitrarily large number of monoexponential decays.
Importantly, there are no explicit convolutions present in \autoref{eq:bigresult}.
Instead, \autoref{eq:bigresult} requires evaluation of complementary error functions; software libraries as diverse as Microsoft Excel and the Scientific Python stack provide this capability using efficient, accurate algorithms.\cite{Zaghloul_Ali_2011, Abrarov_Quine_2018} 
Appendix \ref{Ap3} shows two specific cases of \autoref{eq:bigresult} with the nested summations worked out.

\section{Comparison to experimental results}

We demonstrate the utility and suitability of our strategy of exponentially modified Gaussians for time-resolved florescence in two case studies.
The figure of merit (model error, $\mathcal{E}$) to minimize while fitting is the weighted square residuals,\cite{Grinvald_Steinberg_1974, Knight_Selinger_1971}
\begin{align}
	\mathcal{E} = \sum_i^n w_i \left(\mathcal{F}_{\text{data}}[t_i] - \mathcal{F}_{\text{model}}[t_i] \right)^2,
\end{align} 
where $\mathcal{F}[t_i]$ is the measured fluorescence counts or model prediction at the time point $t_i$.
The weighting factor, $w_i$, is related to the variance, $\sigma_i^2$, at each data point 
\begin{align}
	w_i &=\frac{1}{\sigma_i^2}\left[\frac{1}{n}\sum_i^n\frac{1}{\sigma_i^2}\right]^{-1}, \\
	&= \frac{1}{\mathcal{F}_{\text{data}}[t_i]}\left[\frac{1}{n}\sum_i^n\frac{1}{\mathcal{F}_{\text{data}}[t_i]}\right]^{-1},
\end{align}
in which we have used the fact that photon-counting data without systematic error is well described by a Poisson distribution so the variance and count number, $\mathcal{F}_{\text{data}}$, are equal for a specific time interval. 
To visually access the goodness of fit, we can plot the weighted residuals,
\begin{align}
	\text{weighted residuals} =  \sqrt{w_i} \left(\mathcal{F}_{\text{data}}[t_i] - \mathcal{F}_{\text{model}}[t_i] \right).
\end{align}

\subsection{Fitting wavelength dependent IRF}

\begin{figure*}[!htbp]
	\centering
	\includegraphics[trim=0cm 1cm 2cm 2cm, clip,width=\linewidth]{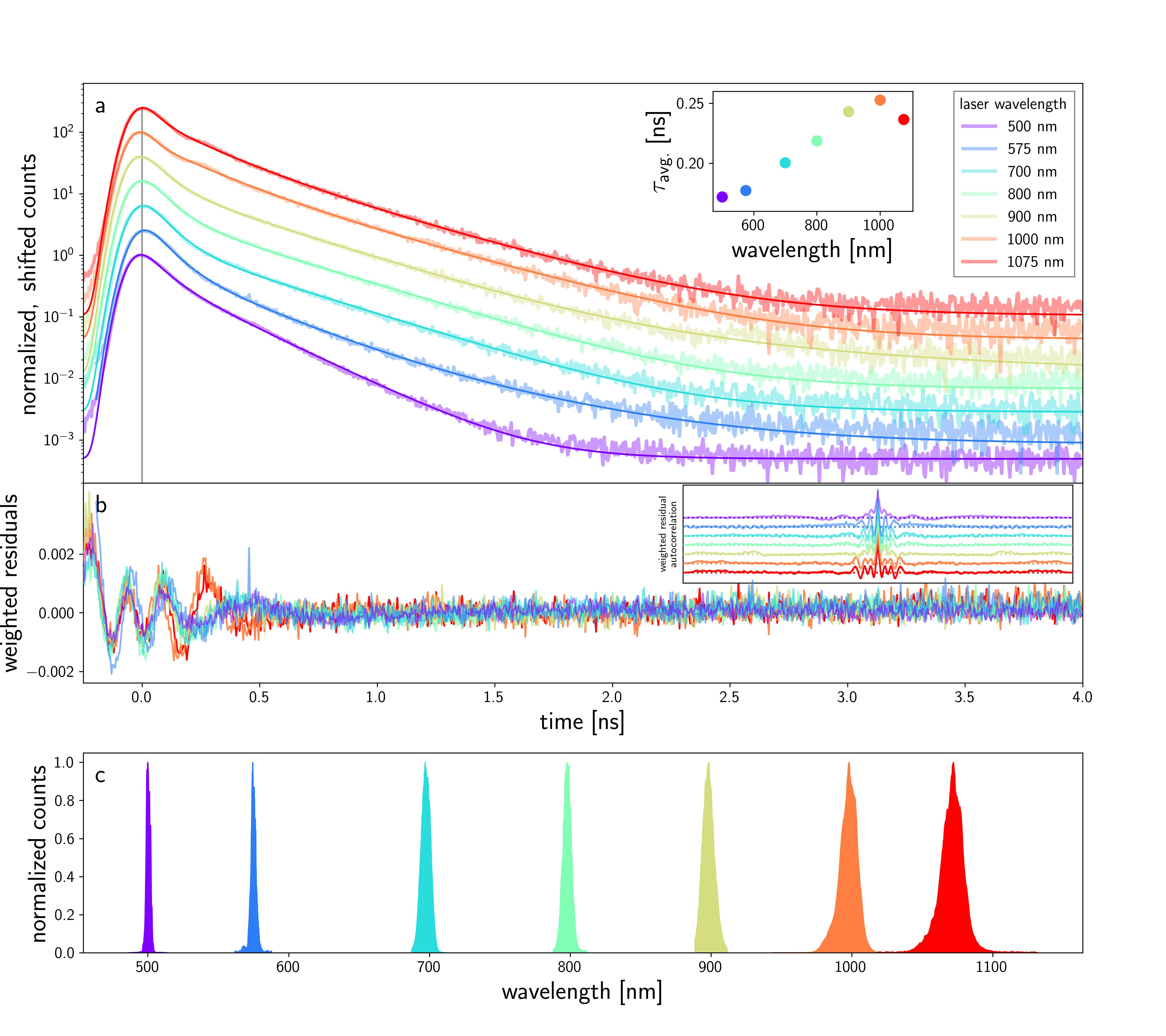}
	\caption{Wavelength dependent SPAD IRF. (a) Temporal response from 500 to 1075 nm with thin, solid lines being the model, $H^{(i)}(x)$.  Pulse profile of excitation laser is shown as gray (note how it is effectively a delta function on the nanosecond scale). Inset shows the intensity weighted average (\autoref{eq:amptau}) of the three decay rates fit to the IRFs. (b) Weighted residuals of fits and their zoomed-in autocorrelation. (c) Color coded spectra of laser used to excite SPAD.
	}\label{fig2:mpd} 
\end{figure*}

In the first study, we characterize the instrument response of a commercial SPAD (PDM from Micro Photon Devices) and counting electronics (PicoQuant HydraHarp 400) by excitation with what is effectively a Dirac delta impulse.
We couple the attenuated output of an ultrafast tuneable Ti:sapphire laser (Coherent Chameleon Discovery with Harmonics Generator, $\sim100$ fs pulse width, 80 MHz repetition rate) into the SPAD. 
By tuning the ultrafast laser from 500 to 1075 nm (\autoref{fig2:mpd}c) and keeping SPAD count rates constant, we recover the wavelength dependent response of the SPAD and associated electronics. 

We fit the IRFs with  $H^{(i)}(x)$ (\autoref{eq:H}) and find that three exponential decays are required for recapitulation of the measurement.
\autoref{fig2:mpd}b shows the weighted residuals and their autocorrelation. 
At delay times before 0.5 ns, the residuals oscillate about zero. 
This indicates that the rising edge of the IRF is not perfectly described by a Gaussian and so \autoref{eq:H} does not exactly capture the instrument response.

The decay profiles shown in \autoref{fig2:mpd}a indicates that as one excites a SPAD with higher wavelength light, the response becomes longer.\cite{VanDenZegel_DeSchryver_1986}
The intensity weighted average decay time (\autoref{fig2:mpd}a inset),\cite{Li_Li_2020}
\begin{align}
	\tau_{\text{avg}} = \frac{\sum_i \frac{\alpha_i}{\lambda_i^2}}{\sum_i \frac{\alpha_i}{\lambda_i}} \label{eq:amptau},
\end{align}
codifies this observation of longer instrument response times for longer wavelength excitation.

\subsection{Fitting wavelength dependent POPOP fluorescence}

\begin{figure*}[!htbp]
	\centering
	\includegraphics[trim=0cm .5cm 2cm 2cm, clip,width=\linewidth]{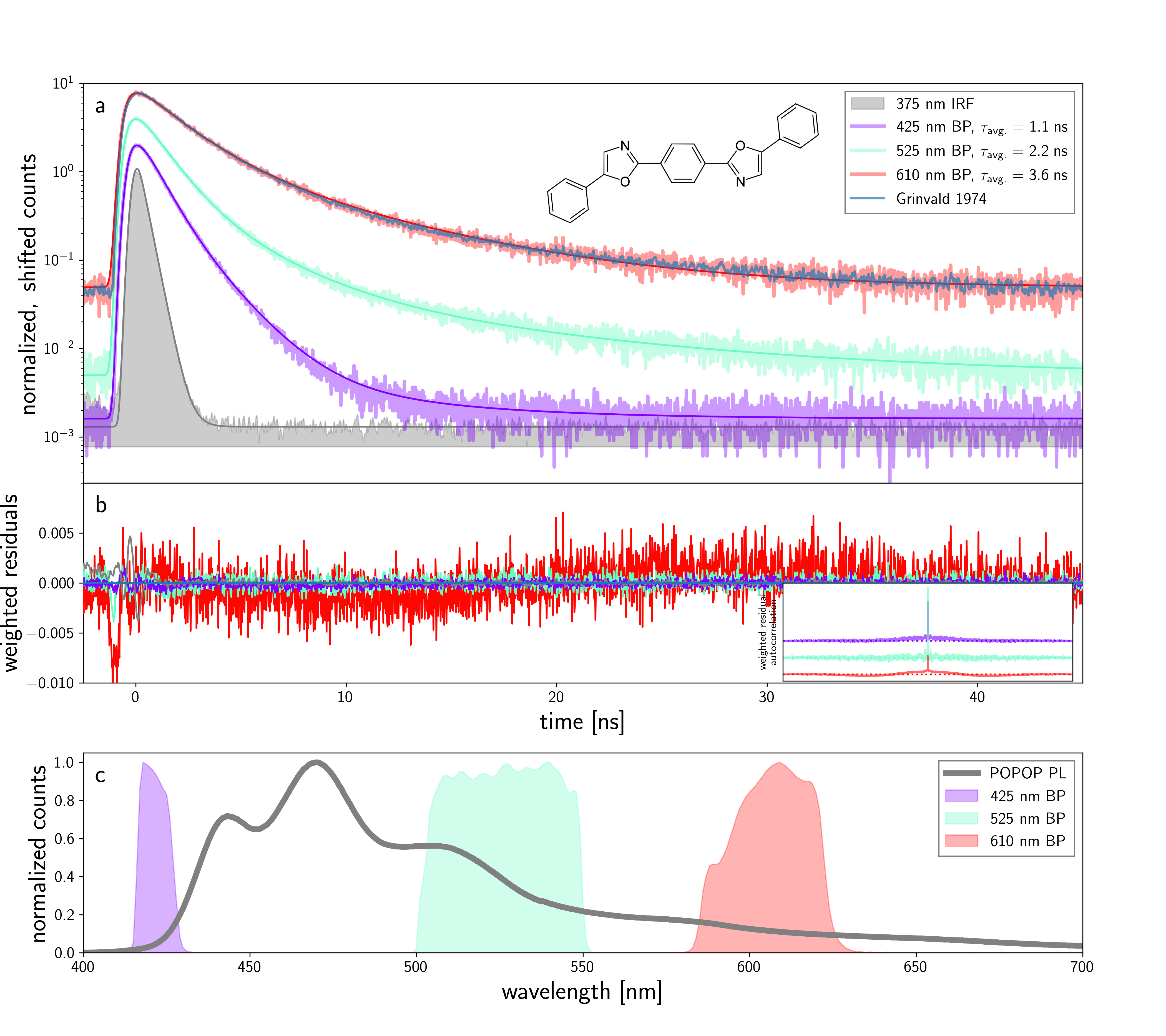}
	\caption{Fluorescence of POPOP excited with a 375 nm pulsed didode laser. 
		(a) Measured and modeled time-resolved florescence. The measured IRF of our SPAD at 375 nm is shown in gray. We recover the following decay rates: for 425 nm, 0.09, 0.14, and 0.16 ns\textsuperscript{-1}; for 525 nm, 0.34, 0.39, and 0.61 ns\textsuperscript{-1}; and finally for 610 nm,  0.59, 0.90, and 1.34 ns\textsuperscript{-1}. A fit of the 610 nm dataset with \textcite{Grinvald_Steinberg_1974}'s iterative method is shown in blue. 
		(b) Weighted residuals of fits and their zoomed-in autocorrelation.
		(c) Fluorescence intensity of POPOP measured through bandpass filters color-coded to traces in (a) along with emission spectra of POPOP which has not been for the wavelength dependent efficiency of our camera and collection optics.
	}\label{fig3:POPOP} 
\end{figure*}

In the second study, we characterize the response of powdered POPOP; 1,4-Bis(5-phenyl-2-oxazolyl)benzene; whose chemical structure is shown in \autoref{fig3:POPOP}a.
POPOP is a scintillator with an absorption onset in the UV-A and bright fluorescence across the visible. 
With an objective we focus the output of a 375 nm pulsed picosecond diode laser (Picoquant, 20 MHz repetition rate) onto powdered POPOP. 
Fluorescence is collected with the same objective, filtered with a 425, 525, or 610 nm bandpass filter, and focused into a photon counting module (PerkinElmer, PicoQuant HydraHarp 400 electronics).
\autoref{fig3:POPOP}c shows a florescence spectrum of POPOP and the throughput of the three bandpass filters used.

We first measure the response function of the detector at 375 nm (\autoref{fig3:POPOP}a).
This IRF is fit in the same way as the IRFs in \autoref{fig2:mpd}a.
Next, we measure the florescence of POPOP over three wavelength bands.
Finally, we use our primary result, \autoref{eq:bigresult}, to fit each time trace (tri-exponential) while holding the IRF parameters constant.
We recover average lifetimes of 1.1, 2.2, and 3.6 ns for the 425, 525, and 610 nm bandpass filters, respectively. 
The model fits all aspects of the time-resolved fluorescence exceedingly well. 
The weighted residuals (\autoref{fig2:mpd}b) show slight systematic deviation at the rising edge of the flouresence near $t=0$.

To compare our model to other methods, we fit the longest wavelength data using the iterative method of \textcite{Grinvald_Steinberg_1974} using the measured IRF. The result of this fit is plotted in blue in \autoref{fig2:mpd}a. 
There is striking agreement between our model and Grinvald's; however, their method better captures the rising edge of the instrument response and does not show a systemic deviation in the early time weighted residuals. 

\section{Comparison of computation time to other methods}

In this Section we show how the computational evaluation time of our EMG formalism compares to other established methods.
Specifically, we look at a quadruple exponential decay convoluted with a Gaussian IRF and compare the evaluation times of (1) a calculation using \autoref{eq:H}, (2) step-wise (iterative) construction of an IRF convoluted decay following the procedure of \textcite{Grinvald_Steinberg_1974}, and (3) numerical convolution of \autoref{eq:exps} with \autoref{eq:g}.
We show the code used to compare these strategies in Algorithms \autoref{alg1} and \autoref{alg2} in the Appendices. 
Base functions are defined in Algorithm \autoref{alg1} using the Scientific Python ecosystem.\cite{Virtanen_Mulbreght_2020, harris2020array, vanderWalt_Varoquaux_2011,vanRossum_2001}
Next, these base functions are used in Algorithm \autoref{alg2} to construct singular functions which each take the \emph{same} input parameters and output a single convoluted decay profile. 
Note, in these constructions it is imperative that the numerical convolution model and \textcite{Grinvald_Steinberg_1974}'s iterative model each be evaluated with an evenly spaced independent (x, delay time) array because the numerical convolution relies on a fast-Fourier-transform and the iterative model uses a trapezoidal (step-sized dependent) sequential integration technique. Our EMG model does not have this restriction, and can be used on datasets with e.g. logarithmically spaced time points. 

\begin{figure}[!htbp]
	\centering
	\includegraphics[trim=0cm .5cm 2cm 2cm, clip,width=0.5\linewidth]{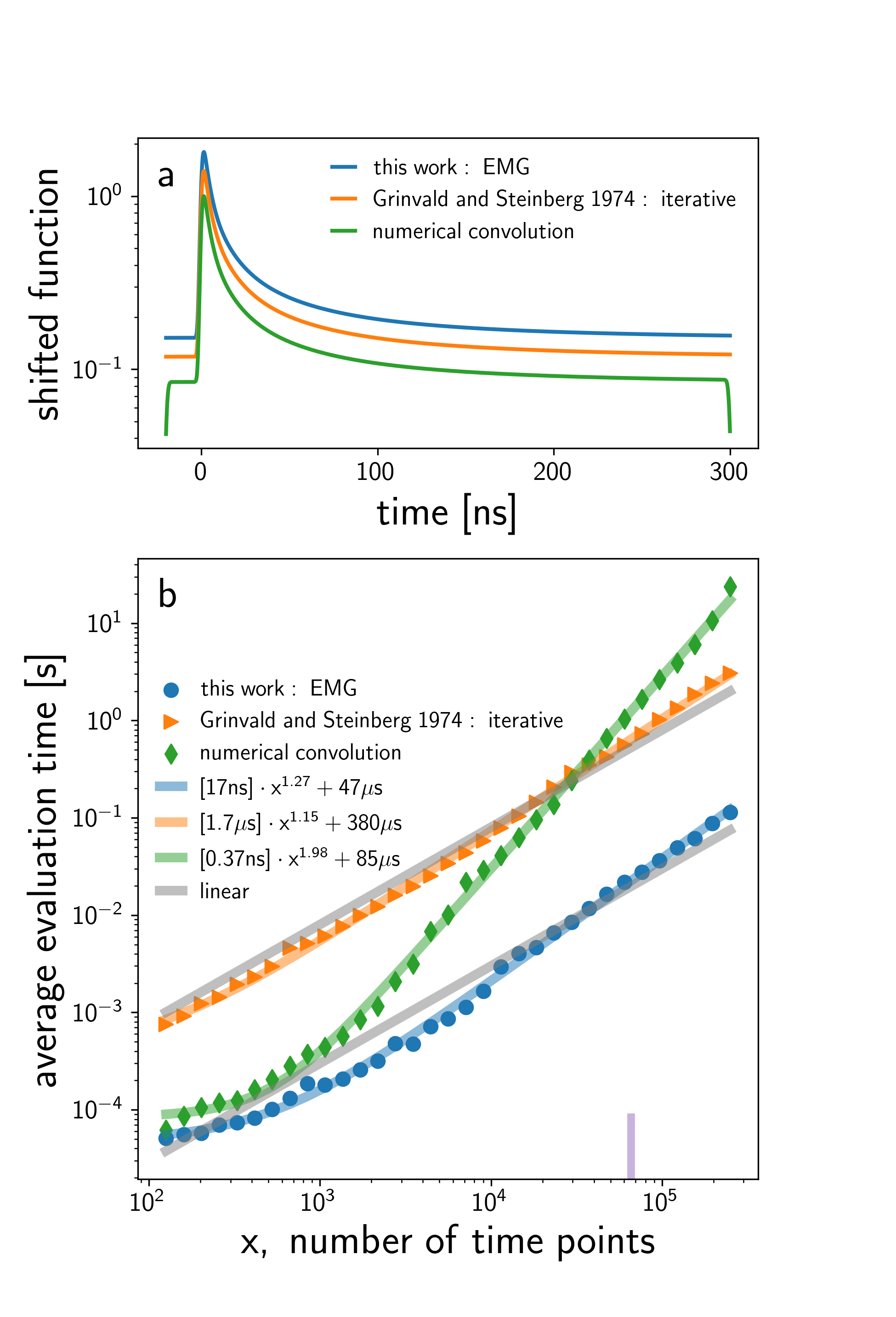}
	\caption{Comparison of \autoref{eq:H}, numerical convolution of \autoref{eq:exps} with \autoref{eq:g}, and the stepwise method of \textcite{Grinvald_Steinberg_1974}. (a) shows that all functions are identical other than edge affects caused by the finite width numerical convolution. (b) shows the average computational evaluation time (Windows 10 running on an Intel Quad Core i7-6700 3.4 GHz processor) as the number of temporal points is increased. The vertical purple bar indicates the number of points (65536 histogram bins) recorded by our HydraHarp 400 time-correlated single photon counting electronics.
	}\label{figevaltime} 
\end{figure}

\autoref{figevaltime}a shows the decay trace from each method. Note, when the traces are not shifted from one-another, they are effectively identical.
The edge affects inherent in numerical convolutions have not been clipped out in order to highlight their existence---our method and \textcite{Grinvald_Steinberg_1974}'s method do not have these edge affects.

We use \texttt{python}'s built in \texttt{timeit} module to measure the \emph{average} execution time of each function as the number of time points in the evaluation are logarithmically increased.
We also logarithmically \emph{decrease} the number of function evaluations from 30000 to 30 as the number of points are increased.
\autoref{figevaltime}b shows the results of this study. 
In all cases, the EMG method is faster than the other two methods. 
On average, the EMG method is $30\times$ faster than \textcite{Grinvald_Steinberg_1974}'s method. This increase in speed likely originates from the fact that our computer program can calculate the exponentially modified Gaussian array in one contiguous batch using an optimized low-level function while the iterative method requires repeated array calls and assignments.\cite{harris2020array, Virtanen_Mulbreght_2020}
 
The EMG method also scales less steeply with number of points ($x^{1.3}$) than the numerical convolution method ($x^{2}$) which will be an important aspect when handling large datasets. 
For datasets with the same number of bins, 65536, as our commercial counting electronics, the EMG method is two orders-of-magnitude faster than direct numerical convolution.

\section{Discussion and Conclusions}

In this work, we derived expressions for fitting the complete transient of time-resolved fluorescence measurements without requiring the use of numerical convolutions. 
We showed that our expressions well reproduce the tailed instrument response of a SPAD.
We then showed that the SPAD IRF could be held constant while a second expression was fit to sample fluorescence.
Our results suggest a clear workflow: 
\begin{enumerate}
	\item Measure instrument response and fit to $$H^{(i)}(x; \mu, \sigma, \{a_i\}, \{\lambda_i\}  ) = \sum_{i}^{m}  a_i h_i(x)$$ using the minimum number of exponential terms to recapitulate the measured response.
	\item Measure sample response and fit to $$\mathcal{M}(x; \{a_j\}, \{\lambda_j\}) = \sum_{i,j}^{m,n} \left(\frac{a_j\lambda_jh_i(t)}{\lambda_j - \lambda_i} + \frac{a_i\lambda_ih_j(t)}{\lambda_i - \lambda_j}\right)$$ using the minimum number of exponential terms to recapitulate the measured response while holding the IRF parameters constant.
\end{enumerate}

While this workflow seems to be robust, there are specific cases where this method should not be applied. 
Firstly, if the material has dynamics which last longer than the period between excitation pulses then there will be some steady-state population which is not accounted for in our model. 
Secondly, if the material has kinetics which are not collectively well described by first order rate laws, then the assumption of exponential decay is poor.
In both of these cases, a more holistic treatment of the fluorescence is needed, indeed there are software packages which can numerically solve the set of pulsed rate equation necessary to correctly describe the system.\cite{Spitha_Wright_2020, Manger_Goldsmith_2017}
Finally, some SPADs are known to exhibit non-ideal behaviors like ``afterpulsing'' which can lead to IRFs which have a train of humps on the tail (note that in \autoref{fig2:mpd}a slight afterpulsing is present around 0.2 ns).\cite{Brown_Rarity_1986,Ziarkash_Ursin_2018}
If a detector's IRF is not well described by a collection of exponentially modified Gaussians which share the same width and center, then our treatment is not adequate. 
Instead, the linearity of convolutions allows one to have multiple weighted, and shifted ($\{\mu_k\}$) versions of the IRF which results in a third summation over $\mu_k$.
However, this level of nested summations is likely excessive and the researcher should instead accomplish  convolutions with their measured, IRF. 

The method presented herein offers a single functional form for the entire rise and decay of time resolved fluorescence data. 
This closed-form may be useful to researchers who do not have access to bespoke software packages for fitting time-resolved fluorescence data.
Perhaps more importantly, because this method has a faster computation time compared to other fitting methods, analytical techniques like fluorescence lifetime imaging microscopy which rely on fitting thousands of decay traces to build up an image will be able to use exponentially modified Gaussians to increase the processing rate of their large streams of data.  

\begin{acknowledgements}
	This work was performed at the Center for Nanoscale Materials, a U.S. Department of Energy Office of Science User Facility, and supported by the U.S. Department of Energy, Office of Science, under Contract No. DE-AC02-06CH11357.
\end{acknowledgements}

\clearpage
\appendix

\begin{widetext}
\section{Showing $h(x)$ to be an EMG} \label{Ap2}
To rewrite $h$ in a form which does not require one to perform numerical convolutions, we must cast the convolution into a complementary error function. We first write out the convolution integral in full
\begin{align}
	h(x; \mu, \sigma, \lambda) &\equiv f(x; \lambda) * g(x; \mu, \sigma), \\
	&=  \left\{\lambda \exp{\left[-\lambda x\right]}\Theta\left[x\right]\right\} * \left\{\frac{1}{\sigma \sqrt{2\pi}} \exp{\left[-\left(\frac{x-\mu}{\sqrt{2}\sigma}\right)^2\right]}\right\}, \\
	&=\frac{\lambda}{\sigma \sqrt{2\pi}} \int_{-\infty}^\infty 
 	\exp{\left[-\lambda t\right]}\Theta\left[t\right]	
	\exp{\left[-\left(\frac{x-t-\mu}{\sqrt{2}\sigma}\right)^2\right]}\textrm{d}t. 
\end{align}
Because of the step function, the integrand is non-zero only when $t>0$ so we can rewrite the bounds of integration
\begin{align}
	h(x; \mu, \sigma, \lambda)	&=\frac{\lambda}{\sigma \sqrt{2\pi}} \int_{0}^\infty 
	\exp{\left[-\lambda t\right]}	
	\exp{\left[-\left(\frac{x-t-\mu}{\sqrt{2}\sigma}\right)^2\right]}\textrm{d}t. \label{eq:usub}
\end{align}
We will perform this integration by substitution.
We define a new variable, $u$,
\begin{align}
	u &\equiv \frac{\mu + \lambda\sigma^2 - x + t}{\sqrt{2}\sigma} \\
	\Longrightarrow \textrm{d}u &= \frac{\textrm{d}t}{\sqrt{2}\sigma} \\
	\Longrightarrow t &= \sqrt{2}\sigma u - \mu -\lambda\sigma^2 + x
\end{align}
Proper substitution of $u$ into \autoref{eq:usub} yields
\begin{align}
	h(x; \mu, \sigma, \lambda)	&=\frac{\sqrt{2}\sigma\lambda}{\sigma \sqrt{2\pi}} \int_{0}^\infty 
	\exp{\left[-\lambda \left(\sqrt{2}\sigma u - \mu -\lambda\sigma^2 + x\right)\right]}	
	\exp{\left[-\left(\frac{x-\left(\sqrt{2}\sigma u - \mu -\lambda\sigma^2 + x\right)-\mu}{\sqrt{2}\sigma}\right)^2\right]}\textrm{d}t, \\
	&=\frac{\lambda}{\sqrt{\pi}} \int_{\frac{\mu+\lambda\sigma^2-x}{\sqrt{2}\sigma}}^\infty 
	\exp{\left[\lambda\mu-\lambda x + \frac{\lambda^2\sigma^2}{2}\right]}	
	\exp{\left[-u^2\right]}	\textrm{d}u.\\
	&=\frac{\lambda}{\sqrt{\pi}}\exp{\left[\lambda\mu-\lambda x + \frac{\lambda^2\sigma^2}{2}\right]} \int_{\frac{\mu+\lambda\sigma^2-x}{\sqrt{2}\sigma}}^\infty 
	\exp{\left[-u^2\right]}	\textrm{d}u. 
\end{align}	
The integral is now in the form of the complementary error function, $\int_{t}^{\infty}\exp{\left[-u^2\right]}\textrm{d}u = \frac{\sqrt{\pi}}{2}\erfc{(t)}$,
\begin{align}
	h(x; \mu, \sigma, \lambda) 	&=\frac{\lambda}{\sqrt{\pi}}\exp{\left[\lambda\mu-\lambda x + \frac{\lambda^2\sigma^2}{2}\right]} \frac{\sqrt{\pi}}{2}\erfc{(t)}, t= \frac{\mu+\lambda\sigma^2-x}{\sqrt{2}\sigma} \\
	 &=  \frac{\lambda}{2}\exp{\left[\frac{\lambda}{2}\left(2\mu + \lambda \sigma^2 - 2x\right)  \right]}  \erfc{\left[\frac{\mu + \lambda\sigma^2 - x}{\sqrt{2}\sigma}\right]}.
\end{align}
This is the desired result, an exponentially modified Gaussian. 

\clearpage
\section{Common cases of Equation 27} \label{Ap3}
In this Appendix we explicitly write out specific cases of \autoref{eq:bigresult}:
\begin{enumerate}
	\item An IRF with one exponential decay $i\in \{1\}$ and two exponential decays in material response $j\in \{\alpha, \beta\}$
	\item An IRF with two exponential decays $i\in \{1,2\}$ and two exponential decays in material response $j\in \{\alpha, \beta\}$
\end{enumerate}

\subsection{One exponential decay in IRF and two decays in material response}

Consider the case of an IRF which has one exponential decay, $i\in \{1\}$ and a material response with two exponential decays, $j\in \{\alpha, \beta\}$.
\begin{align}
	\mathcal{M}_{\{1\},\{\alpha, \beta\}}(x) = {}& h_1(x)\left(\frac{a_\alpha\lambda_\alpha}{\lambda_\alpha - \lambda_1} + \frac{a_\beta\lambda_\beta}{\lambda_\beta - \lambda_1}\right)  +
	a_1\lambda_1 \left\{\frac{h_\alpha(x)}{\lambda_1-\lambda_\alpha} + \frac{h_\beta(x)}{\lambda_1-\lambda_\beta}  \right\} \\
	h_1(x) = {}& \frac{\lambda_1}{2}\exp{\left[\frac{\lambda_1}{2}\left(2\mu + \lambda_1 \sigma^2 - 2x\right)  \right]} \erfc{\left[\frac{\mu + \lambda_1\sigma^2 - x}{\sqrt{2}\sigma}\right]} \\
	h_\alpha(x) = {}& \frac{\lambda_\alpha}{2}\exp{\left[\frac{\lambda_\alpha}{2}\left(2\mu + \lambda_\alpha \sigma^2 - 2x\right)  \right]} \erfc{\left[\frac{\mu + \lambda_\alpha\sigma^2 - x}{\sqrt{2}\sigma}\right]} \\
	h_\beta(x) = {}& \frac{\lambda_\beta}{2}\exp{\left[\frac{\lambda_\beta}{2}\left(2\mu + \lambda_\beta \sigma^2 - 2x\right)  \right]} \erfc{\left[\frac{\mu + \lambda_\beta\sigma^2 - x}{\sqrt{2}\sigma}\right]}
\end{align} 
First one would fit the measured IRF using the parameters $\{\sigma, \mu, a_1, \lambda_1\}$. 
Then these parameters would be held constant and the full response would be fit using the parameters $\{a_\alpha, a_\beta, \lambda_\alpha, \lambda_\beta\}$. 

\subsection{Two exponential decays in both IRF and material response}

Consider the case of an IRF which has two exponential decays, $i\in \{1,2\}$ and a material response also with two exponential decays, $j\in \{\alpha, \beta\}$.
\begin{align}
	\begin{split}
		\mathcal{M}_{\{1,2\},\{\alpha, \beta\}}(x) = {}& + h_1(x)\left(\frac{a_\alpha\lambda_\alpha}{\lambda_\alpha - \lambda_1} + \frac{a_\beta\lambda_\beta}{\lambda_\beta - \lambda_1}\right) +
		h_2(x)\left(\frac{a_\alpha\lambda_\alpha}{\lambda_\alpha - \lambda_2} + \frac{a_\beta\lambda_\beta}{\lambda_\beta - \lambda_2}\right)  \\
		&+ h_\alpha(x)\left(\frac{a_1\lambda_1}{\lambda_1 - \lambda_\alpha} + \frac{a_2\lambda_2}{\lambda_2 - \lambda_\alpha}\right) +
		h_\beta(x)\left(\frac{a_1\lambda_1}{\lambda_1 - \lambda_\beta} + \frac{a_2\lambda_2}{\lambda_2 - \lambda_\beta}\right)
	\end{split} \\
	h_1(x) = {}& \frac{\lambda_1}{2}\exp{\left[\frac{\lambda_1}{2}\left(2\mu + \lambda_1 \sigma^2 - 2x\right)  \right]} \erfc{\left[\frac{\mu + \lambda_1\sigma^2 - x}{\sqrt{2}\sigma}\right]} \\
	h_2(x) = {}& \frac{\lambda_2}{2}\exp{\left[\frac{\lambda_2}{2}\left(2\mu + \lambda_2 \sigma^2 - 2x\right)  \right]} \erfc{\left[\frac{\mu + \lambda_2\sigma^2 - x}{\sqrt{2}\sigma}\right]} \\
	h_\alpha(x) = {}& \frac{\lambda_\alpha}{2}\exp{\left[\frac{\lambda_\alpha}{2}\left(2\mu + \lambda_\alpha \sigma^2 - 2x\right)  \right]} \erfc{\left[\frac{\mu + \lambda_\alpha\sigma^2 - x}{\sqrt{2}\sigma}\right]} \\
	h_\beta(x) = {}& \frac{\lambda_\beta}{2}\exp{\left[\frac{\lambda_\beta}{2}\left(2\mu + \lambda_\beta \sigma^2 - 2x\right)  \right]} \erfc{\left[\frac{\mu + \lambda_\beta\sigma^2 - x}{\sqrt{2}\sigma}\right]}
\end{align} 
First one would fit the measured IRF using the parameters $\{\sigma, \mu, a_1, a_2, \lambda_1, \lambda_2\}$. 
Then these parameters would be held constant and the full response would be fit using the parameters $\{a_\alpha, a_\beta, \lambda_\alpha, \lambda_\beta\}$. 

\clearpage		
\section{Function evaluation time}

In this Appendix we show the code used to determine the computational evaluation time of our exponentially modified Gaussian formalism versus other established methods.
Specifically, we look at a quadruple exponential decay convoluted with a Gaussian IRF. 
We compare the evaluation times of:
\begin{enumerate}
	\item Analytical calculation using \autoref{eq:H}.
	\item Numerical convolution of \autoref{eq:exps} with \autoref{eq:g}.
	\item Stepwise (iterative) construction of an IRF convoluted decay following the procedure of \textcite{Grinvald_Steinberg_1974}.
\end{enumerate}

\begin{figure}[!htbp]
	\begin{algorithm}[H]
		\caption{\texttt{python 3.8} construction of exponential, Gaussian, and exponentially modified Gaussian functions.}
		\label{alg1}
\begin{verbatim}
import numpy as np
import timeit
from scipy import special
	
def f(x, l, a=1):
    # exponential decay with heaviside
    out = a * l * np.exp(-l * x)
    out[x < 0] = 0
    return out    
	
def g(x, mu, s):
    # Gaussian
    arg = -1 * ((x - mu) / (np.sqrt(2) * s)) ** 2
    return np.exp(arg) / (s * np.sqrt(2 * np.pi))
	
def h(x, mu, s, l, a=1):
    # exponentially modified Gaussian (EMG)
    erfc = special.erfc
    arg1 = l / 2 * (2 * mu + l * s**2 - 2 * x)
    arg2 = (mu + l * s**2 - x) / (np.sqrt(2) * s)
    return a * l / 2 * np.exp(arg1) * erfc(arg2)
	
def F(x, li, ai):
    # summation of exponentials with Heaviside
    # assumes x, li, and ai are all 1D. li and ai have same size
    out = ai[None, :] * li[None, :] * np.exp(-li[None, :] * x[:, None])
    out = np.sum(out, axis=-1)
    out[x < 0] = 0
    return out
	
def H(x, mu, s, li, ai):
    # summation of EMGs with only one Gaussian convolved.
    # assumes x, li, and ai are all 1D. li and ai have same size
    out = h(x[:, None], mu, s, li[None, :], ai[None, :])
    return np.sum(out, axis=-1)
\end{verbatim}
	\end{algorithm}
\end{figure}

\begin{figure}[!htbp]
	\begin{algorithm}[H]
		\caption{\texttt{python 3.8} construction of quad-exponential functions and evaluation time assessment.}
		\label{alg2}
\begin{verbatim}
def quadexp_unpack(p):
    # unpack parameter array
    mu, s = p[0], p[1]
    li = p[2:6]
    ai = p[6:10]
    off = p[10]
    return mu, s, li, ai, off
	
def quadexp_EMG_model(p, x):
    # exponentially modified Gaussian
    mu, s, li, ai, off = quadexp_unpack(p)
    return H(x, mu, s, li, ai) + off
	
def quadexp_convolution_model(p, x):
    # direct numerical convolution of Gaussian with exponentials
    mu, s, li, ai, off = quadexp_unpack(p)
    IRF = g(x - x.mean(), mu, s)
    IRF /= IRF.sum()  # area normalization
    decay = F(x, li, ai) + off
    return np.convolve(decay, IRF, mode="same")
	
def quadexp_iterative_model(p, x):
    # Iterative model of Grinvald and Steinberg 1974
    # See equation A-5 of DOI: 10.1016/0003-2697(74)90312-1
    outs = np.zeros((x.size, 4))
    epsilon = x[1] - x[0]
    mu, s, li, ai, off = quadexp_unpack(p)
    IRF = g(x, mu, s)
    exp = np.exp(-epsilon * li)
    epsilon_a = epsilon * ai * li
    for i in range(x.size - 1):
        outs[i + 1] = (outs[i] + 0.5 * epsilon_a * IRF[i]) * exp + 0.5 * epsilon_a * IRF[i + 1]
    return np.sum(outs, axis=-1) + off
	
def time_function_evaluation(xsize, numevals):
    # returns array with average evaluation time of 3 different models
    # ['EMG', 'numerical convolution', 'iterative']
    x = np.linspace(-20, 300, xsize)
    out = np.zeros(3)
    p = np.array([0, 1, 1 / 3, 1 / 10, 1 / 30, 1 / 100, 0.3, 0.4, 0.5, 0.6, 0.01])
    out[0] = timeit.timeit(lambda: quadexp_EMG_model(p, x), number=numevals)
    out[1] = timeit.timeit(lambda: quadexp_convolution_model(p, x), number=numevals)
    out[2] = timeit.timeit(lambda: quadexp_iterative_model(p, x), number=numevals)
    return out / numevals
	
# average evaluation time in seconds
#['EMG', 'numerical convolution', 'iterative']
print(time_function_evaluation(xsize=65536, numevals=10))
>>> [0.01594715 0.50203172 0.39133182]
\end{verbatim}
	\end{algorithm}
\end{figure}

\clearpage
\end{widetext}
\bibliography{database}

\begin{thebibliography}{32}%
\makeatletter
\providecommand \@ifxundefined [1]{%
 \@ifx{#1\undefined}
}%
\providecommand \@ifnum [1]{%
 \ifnum #1\expandafter \@firstoftwo
 \else \expandafter \@secondoftwo
 \fi
}%
\providecommand \@ifx [1]{%
 \ifx #1\expandafter \@firstoftwo
 \else \expandafter \@secondoftwo
 \fi
}%
\providecommand \natexlab [1]{#1}%
\providecommand \enquote  [1]{``#1''}%
\providecommand \bibnamefont  [1]{#1}%
\providecommand \bibfnamefont [1]{#1}%
\providecommand \citenamefont [1]{#1}%
\providecommand \href@noop [0]{\@secondoftwo}%
\providecommand \href [0]{\begingroup \@sanitize@url \@href}%
\providecommand \@href[1]{\@@startlink{#1}\@@href}%
\providecommand \@@href[1]{\endgroup#1\@@endlink}%
\providecommand \@sanitize@url [0]{\catcode `\\12\catcode `\$12\catcode
  `\&12\catcode `\#12\catcode `\^12\catcode `\_12\catcode `\%12\relax}%
\providecommand \@@startlink[1]{}%
\providecommand \@@endlink[0]{}%
\providecommand \url  [0]{\begingroup\@sanitize@url \@url }%
\providecommand \@url [1]{\endgroup\@href {#1}{\urlprefix }}%
\providecommand \urlprefix  [0]{URL }%
\providecommand \Eprint [0]{\href }%
\providecommand \doibase [0]{https://doi.org/}%
\providecommand \selectlanguage [0]{\@gobble}%
\providecommand \bibinfo  [0]{\@secondoftwo}%
\providecommand \bibfield  [0]{\@secondoftwo}%
\providecommand \translation [1]{[#1]}%
\providecommand \BibitemOpen [0]{}%
\providecommand \bibitemStop [0]{}%
\providecommand \bibitemNoStop [0]{.\EOS\space}%
\providecommand \EOS [0]{\spacefactor3000\relax}%
\providecommand \BibitemShut  [1]{\csname bibitem#1\endcsname}%
\let\auto@bib@innerbib\@empty
%</preamble>
\bibitem [{\citenamefont {Fu}\ \emph {et~al.}(2017)\citenamefont {Fu},
  \citenamefont {Rea}, \citenamefont {Chen}, \citenamefont {Morrow},
  \citenamefont {Hautzinger}, \citenamefont {Zhao}, \citenamefont {Pan},
  \citenamefont {Manger}, \citenamefont {Wright}, \citenamefont {Goldsmith},\
  and\ \citenamefont {Jin}}]{Fu_Jin_2017}%
  \BibitemOpen
  \bibfield  {author} {\bibinfo {author} {\bibfnamefont {Y.}~\bibnamefont
  {Fu}}, \bibinfo {author} {\bibfnamefont {M.~T.}\ \bibnamefont {Rea}},
  \bibinfo {author} {\bibfnamefont {J.}~\bibnamefont {Chen}}, \bibinfo {author}
  {\bibfnamefont {D.~J.}\ \bibnamefont {Morrow}}, \bibinfo {author}
  {\bibfnamefont {M.~P.}\ \bibnamefont {Hautzinger}}, \bibinfo {author}
  {\bibfnamefont {Y.}~\bibnamefont {Zhao}}, \bibinfo {author} {\bibfnamefont
  {D.}~\bibnamefont {Pan}}, \bibinfo {author} {\bibfnamefont {L.~H.}\
  \bibnamefont {Manger}}, \bibinfo {author} {\bibfnamefont {J.~C.}\
  \bibnamefont {Wright}}, \bibinfo {author} {\bibfnamefont {R.~H.}\
  \bibnamefont {Goldsmith}},\ and\ \bibinfo {author} {\bibfnamefont
  {S.}~\bibnamefont {Jin}},\ }\bibfield  {title} {\bibinfo {title} {Selective
  stabilization and photophysical properties of metastable perovskite
  polymorphs of {CsPbI}$_3$ in thin films},\ }\href
  {https://doi.org/10.1021/acs.chemmater.7b02948} {\bibfield  {journal}
  {\bibinfo  {journal} {Chem. Mater.}\ }\textbf {\bibinfo {volume} {29}},\
  \bibinfo {pages} {8385} (\bibinfo {year} {2017})}\BibitemShut {NoStop}%
\bibitem [{\citenamefont {Goldsmith}\ and\ \citenamefont
  {Moerner}(2010)}]{Goldsmith_Moerner_2010}%
  \BibitemOpen
  \bibfield  {author} {\bibinfo {author} {\bibfnamefont {R.~H.}\ \bibnamefont
  {Goldsmith}}\ and\ \bibinfo {author} {\bibfnamefont {W.~E.}\ \bibnamefont
  {Moerner}},\ }\bibfield  {title} {\bibinfo {title} {Watching conformational-
  and photodynamics of single fluorescent proteins in solution},\ }\href
  {https://doi.org/10.1038/nchem.545} {\bibfield  {journal} {\bibinfo
  {journal} {Nat. Chem.}\ }\textbf {\bibinfo {volume} {2}},\ \bibinfo {pages}
  {179} (\bibinfo {year} {2010})}\BibitemShut {NoStop}%
\bibitem [{\citenamefont {Elson}\ \emph {et~al.}(2004)\citenamefont {Elson},
  \citenamefont {Requejo-Isidro}, \citenamefont {Munro}, \citenamefont
  {Reavell}, \citenamefont {Siegel}, \citenamefont {Suhling}, \citenamefont
  {Tadrous}, \citenamefont {Benninger}, \citenamefont {Lanigan}, \citenamefont
  {McGinty}, \citenamefont {Talbot}, \citenamefont {Treanor}, \citenamefont
  {Webb}, \citenamefont {Sandison}, \citenamefont {Wallace}, \citenamefont
  {Davis}, \citenamefont {Lever}, \citenamefont {Neil}, \citenamefont
  {Phillips}, \citenamefont {Stamp},\ and\ \citenamefont
  {French}}]{Elson_French_2004}%
  \BibitemOpen
  \bibfield  {author} {\bibinfo {author} {\bibfnamefont {D.}~\bibnamefont
  {Elson}}, \bibinfo {author} {\bibfnamefont {J.}~\bibnamefont
  {Requejo-Isidro}}, \bibinfo {author} {\bibfnamefont {I.}~\bibnamefont
  {Munro}}, \bibinfo {author} {\bibfnamefont {F.}~\bibnamefont {Reavell}},
  \bibinfo {author} {\bibfnamefont {J.}~\bibnamefont {Siegel}}, \bibinfo
  {author} {\bibfnamefont {K.}~\bibnamefont {Suhling}}, \bibinfo {author}
  {\bibfnamefont {P.}~\bibnamefont {Tadrous}}, \bibinfo {author} {\bibfnamefont
  {R.}~\bibnamefont {Benninger}}, \bibinfo {author} {\bibfnamefont
  {P.}~\bibnamefont {Lanigan}}, \bibinfo {author} {\bibfnamefont
  {J.}~\bibnamefont {McGinty}}, \bibinfo {author} {\bibfnamefont
  {C.}~\bibnamefont {Talbot}}, \bibinfo {author} {\bibfnamefont
  {B.}~\bibnamefont {Treanor}}, \bibinfo {author} {\bibfnamefont
  {S.}~\bibnamefont {Webb}}, \bibinfo {author} {\bibfnamefont {A.}~\bibnamefont
  {Sandison}}, \bibinfo {author} {\bibfnamefont {A.}~\bibnamefont {Wallace}},
  \bibinfo {author} {\bibfnamefont {D.}~\bibnamefont {Davis}}, \bibinfo
  {author} {\bibfnamefont {J.}~\bibnamefont {Lever}}, \bibinfo {author}
  {\bibfnamefont {M.}~\bibnamefont {Neil}}, \bibinfo {author} {\bibfnamefont
  {D.}~\bibnamefont {Phillips}}, \bibinfo {author} {\bibfnamefont
  {G.}~\bibnamefont {Stamp}},\ and\ \bibinfo {author} {\bibfnamefont
  {P.}~\bibnamefont {French}},\ }\bibfield  {title} {\bibinfo {title}
  {Time-domain fluorescence lifetime imaging applied to biological tissue},\
  }\href {https://doi.org/10.1039/b316456j} {\bibfield  {journal} {\bibinfo
  {journal} {Photochem. Photobiol. Sci.}\ }\textbf {\bibinfo {volume} {3}},\
  \bibinfo {pages} {795} (\bibinfo {year} {2004})}\BibitemShut {NoStop}%
\bibitem [{\citenamefont {Bastiaens}\ and\ \citenamefont
  {Squire}(1999)}]{Bastiaens_Squire_1999}%
  \BibitemOpen
  \bibfield  {author} {\bibinfo {author} {\bibfnamefont {P.~I.}\ \bibnamefont
  {Bastiaens}}\ and\ \bibinfo {author} {\bibfnamefont {A.}~\bibnamefont
  {Squire}},\ }\bibfield  {title} {\bibinfo {title} {Fluorescence lifetime
  imaging microscopy: spatial resolution of biochemical processes in the
  cell},\ }\href {https://doi.org/10.1016/S0962-8924(98)01410-X} {\bibfield
  {journal} {\bibinfo  {journal} {Trends Cell Biol.}\ }\textbf {\bibinfo
  {volume} {9}},\ \bibinfo {pages} {48} (\bibinfo {year} {1999})}\BibitemShut
  {NoStop}%
\bibitem [{\citenamefont {Gorpas}\ \emph {et~al.}(2019)\citenamefont {Gorpas},
  \citenamefont {Phipps}, \citenamefont {Bec}, \citenamefont {Ma},
  \citenamefont {Dochow}, \citenamefont {Yankelevich}, \citenamefont {Sorger},
  \citenamefont {Popp}, \citenamefont {Bewley}, \citenamefont
  {Gandour-Edwards}, \citenamefont {Marcu},\ and\ \citenamefont
  {Farwell}}]{Gorpas_Farwell_2019}%
  \BibitemOpen
  \bibfield  {author} {\bibinfo {author} {\bibfnamefont {D.}~\bibnamefont
  {Gorpas}}, \bibinfo {author} {\bibfnamefont {J.}~\bibnamefont {Phipps}},
  \bibinfo {author} {\bibfnamefont {J.}~\bibnamefont {Bec}}, \bibinfo {author}
  {\bibfnamefont {D.}~\bibnamefont {Ma}}, \bibinfo {author} {\bibfnamefont
  {S.}~\bibnamefont {Dochow}}, \bibinfo {author} {\bibfnamefont
  {D.}~\bibnamefont {Yankelevich}}, \bibinfo {author} {\bibfnamefont
  {J.}~\bibnamefont {Sorger}}, \bibinfo {author} {\bibfnamefont
  {J.}~\bibnamefont {Popp}}, \bibinfo {author} {\bibfnamefont {A.}~\bibnamefont
  {Bewley}}, \bibinfo {author} {\bibfnamefont {R.}~\bibnamefont
  {Gandour-Edwards}}, \bibinfo {author} {\bibfnamefont {L.}~\bibnamefont
  {Marcu}},\ and\ \bibinfo {author} {\bibfnamefont {D.~G.}\ \bibnamefont
  {Farwell}},\ }\bibfield  {title} {\bibinfo {title} {Autofluorescence lifetime
  augmented reality as a means for real-time robotic surgery guidance in human
  patients},\ }\href {https://doi.org/10.1038/s41598-018-37237-8} {\bibfield
  {journal} {\bibinfo  {journal} {Sci. Rep.}\ }\textbf {\bibinfo {volume}
  {9}},\ \bibinfo {pages} {1187} (\bibinfo {year} {2019})}\BibitemShut
  {NoStop}%
\bibitem [{\citenamefont {Comelli}\ \emph {et~al.}(2004)\citenamefont
  {Comelli}, \citenamefont {D'Andrea}, \citenamefont {Valentini}, \citenamefont
  {Cubeddu}, \citenamefont {Colombo},\ and\ \citenamefont
  {Toniolo}}]{Comelli_Toniolo_2004}%
  \BibitemOpen
  \bibfield  {author} {\bibinfo {author} {\bibfnamefont {D.}~\bibnamefont
  {Comelli}}, \bibinfo {author} {\bibfnamefont {C.}~\bibnamefont {D'Andrea}},
  \bibinfo {author} {\bibfnamefont {G.}~\bibnamefont {Valentini}}, \bibinfo
  {author} {\bibfnamefont {R.}~\bibnamefont {Cubeddu}}, \bibinfo {author}
  {\bibfnamefont {C.}~\bibnamefont {Colombo}},\ and\ \bibinfo {author}
  {\bibfnamefont {L.}~\bibnamefont {Toniolo}},\ }\bibfield  {title} {\bibinfo
  {title} {Fluorescence lifetime imaging and spectroscopy as tools for
  nondestructive analysis of works of art},\ }\href
  {https://doi.org/10.1364/ao.43.002175} {\bibfield  {journal} {\bibinfo
  {journal} {Appl. Opt.}\ }\textbf {\bibinfo {volume} {43}},\ \bibinfo {pages}
  {2175} (\bibinfo {year} {2004})}\BibitemShut {NoStop}%
\bibitem [{\citenamefont {Knight}\ and\ \citenamefont
  {Selinger}(1971)}]{Knight_Selinger_1971}%
  \BibitemOpen
  \bibfield  {author} {\bibinfo {author} {\bibfnamefont {A.}~\bibnamefont
  {Knight}}\ and\ \bibinfo {author} {\bibfnamefont {B.}~\bibnamefont
  {Selinger}},\ }\bibfield  {title} {\bibinfo {title} {The deconvolution of
  fluorescence decay curves},\ }\href
  {https://doi.org/10.1016/0584-8539(71)80073-9} {\bibfield  {journal}
  {\bibinfo  {journal} {Spectrochim. Acta - A: Mol. Biomol.}\ }\textbf
  {\bibinfo {volume} {27}},\ \bibinfo {pages} {1223} (\bibinfo {year}
  {1971})}\BibitemShut {NoStop}%
\bibitem [{\citenamefont {Becker}(2012)}]{Becker_2012}%
  \BibitemOpen
  \bibfield  {author} {\bibinfo {author} {\bibfnamefont {W.}~\bibnamefont
  {Becker}},\ }\bibfield  {title} {\bibinfo {title} {Fluorescence lifetime
  imaging - techniques and applications},\ }\href
  {https://doi.org/10.1111/j.1365-2818.2012.03618.x} {\bibfield  {journal}
  {\bibinfo  {journal} {J. Microsc.}\ }\textbf {\bibinfo {volume} {247}},\
  \bibinfo {pages} {119} (\bibinfo {year} {2012})}\BibitemShut {NoStop}%
\bibitem [{\citenamefont {Warren}\ \emph {et~al.}(2013)\citenamefont {Warren},
  \citenamefont {Margineanu}, \citenamefont {Alibhai}, \citenamefont {Kelly},
  \citenamefont {Talbot}, \citenamefont {Alexandrov}, \citenamefont {Munro},
  \citenamefont {Katan}, \citenamefont {Dunsby},\ and\ \citenamefont
  {French}}]{Warren_Degtyar_2013}%
  \BibitemOpen
  \bibfield  {author} {\bibinfo {author} {\bibfnamefont {S.~C.}\ \bibnamefont
  {Warren}}, \bibinfo {author} {\bibfnamefont {A.}~\bibnamefont {Margineanu}},
  \bibinfo {author} {\bibfnamefont {D.}~\bibnamefont {Alibhai}}, \bibinfo
  {author} {\bibfnamefont {D.~J.}\ \bibnamefont {Kelly}}, \bibinfo {author}
  {\bibfnamefont {C.}~\bibnamefont {Talbot}}, \bibinfo {author} {\bibfnamefont
  {Y.}~\bibnamefont {Alexandrov}}, \bibinfo {author} {\bibfnamefont
  {I.}~\bibnamefont {Munro}}, \bibinfo {author} {\bibfnamefont
  {M.}~\bibnamefont {Katan}}, \bibinfo {author} {\bibfnamefont
  {C.}~\bibnamefont {Dunsby}},\ and\ \bibinfo {author} {\bibfnamefont
  {P.~M.~W.}\ \bibnamefont {French}},\ }\bibfield  {title} {\bibinfo {title}
  {Rapid global fitting of large fluorescence lifetime imaging microscopy
  datasets},\ }\href {https://doi.org/10.1371/journal.pone.0070687} {\bibfield
  {journal} {\bibinfo  {journal} {PLoS ONE}\ }\textbf {\bibinfo {volume} {8}},\
  \bibinfo {pages} {e70687} (\bibinfo {year} {2013})}\BibitemShut {NoStop}%
\bibitem [{\citenamefont {Jeansonne}\ and\ \citenamefont
  {Foley}(1991)}]{Jeansonne_Foley_1991}%
  \BibitemOpen
  \bibfield  {author} {\bibinfo {author} {\bibfnamefont {M.~S.}\ \bibnamefont
  {Jeansonne}}\ and\ \bibinfo {author} {\bibfnamefont {J.~P.}\ \bibnamefont
  {Foley}},\ }\bibfield  {title} {\bibinfo {title} {Review of the exponentially
  modified gaussian ({EMG}) function since 1983},\ }\href
  {https://doi.org/10.1093/chromsci/29.6.258} {\bibfield  {journal} {\bibinfo
  {journal} {J. Chromatogr. Sci.}\ }\textbf {\bibinfo {volume} {29}},\ \bibinfo
  {pages} {258} (\bibinfo {year} {1991})}\BibitemShut {NoStop}%
\bibitem [{\citenamefont {Kalambet}\ \emph {et~al.}(2011)\citenamefont
  {Kalambet}, \citenamefont {Kozmin}, \citenamefont {Mikhailova}, \citenamefont
  {Nagaev},\ and\ \citenamefont {Tikhonov}}]{Kalambet_Tikhonov_2011}%
  \BibitemOpen
  \bibfield  {author} {\bibinfo {author} {\bibfnamefont {Y.}~\bibnamefont
  {Kalambet}}, \bibinfo {author} {\bibfnamefont {Y.}~\bibnamefont {Kozmin}},
  \bibinfo {author} {\bibfnamefont {K.}~\bibnamefont {Mikhailova}}, \bibinfo
  {author} {\bibfnamefont {I.}~\bibnamefont {Nagaev}},\ and\ \bibinfo {author}
  {\bibfnamefont {P.}~\bibnamefont {Tikhonov}},\ }\bibfield  {title} {\bibinfo
  {title} {Reconstruction of chromatographic peaks using the exponentially
  modified gaussian function},\ }\href {https://doi.org/10.1002/cem.1343}
  {\bibfield  {journal} {\bibinfo  {journal} {J. Chemometrics}\ }\textbf
  {\bibinfo {volume} {25}},\ \bibinfo {pages} {352} (\bibinfo {year}
  {2011})}\BibitemShut {NoStop}%
\bibitem [{\citenamefont {Purushothaman}\ \emph {et~al.}(2017)\citenamefont
  {Purushothaman}, \citenamefont {Andr{\'{e}}s}, \citenamefont {Bergmann},
  \citenamefont {Dickel}, \citenamefont {Ebert}, \citenamefont {Geissel},
  \citenamefont {Hornung}, \citenamefont {Pla{\ss}}, \citenamefont {Rappold},
  \citenamefont {Scheidenberger}, \citenamefont {Tanaka},\ and\ \citenamefont
  {Yavor}}]{Purushothaman_Yavor_2017}%
  \BibitemOpen
  \bibfield  {author} {\bibinfo {author} {\bibfnamefont {S.}~\bibnamefont
  {Purushothaman}}, \bibinfo {author} {\bibfnamefont {S.~A.~S.}\ \bibnamefont
  {Andr{\'{e}}s}}, \bibinfo {author} {\bibfnamefont {J.}~\bibnamefont
  {Bergmann}}, \bibinfo {author} {\bibfnamefont {T.}~\bibnamefont {Dickel}},
  \bibinfo {author} {\bibfnamefont {J.}~\bibnamefont {Ebert}}, \bibinfo
  {author} {\bibfnamefont {H.}~\bibnamefont {Geissel}}, \bibinfo {author}
  {\bibfnamefont {C.}~\bibnamefont {Hornung}}, \bibinfo {author} {\bibfnamefont
  {W.}~\bibnamefont {Pla{\ss}}}, \bibinfo {author} {\bibfnamefont
  {C.}~\bibnamefont {Rappold}}, \bibinfo {author} {\bibfnamefont
  {C.}~\bibnamefont {Scheidenberger}}, \bibinfo {author} {\bibfnamefont
  {Y.}~\bibnamefont {Tanaka}},\ and\ \bibinfo {author} {\bibfnamefont
  {M.}~\bibnamefont {Yavor}},\ }\bibfield  {title} {\bibinfo {title}
  {Hyper-{EMG}: A new probability distribution function composed of
  exponentially modified gaussian distributions to analyze asymmetric peak
  shapes in high-resolution time-of-flight mass spectrometry},\ }\href
  {https://doi.org/10.1016/j.ijms.2017.07.014} {\bibfield  {journal} {\bibinfo
  {journal} {Int. J. Mass. Spectrom.}\ }\textbf {\bibinfo {volume} {421}},\
  \bibinfo {pages} {245} (\bibinfo {year} {2017})}\BibitemShut {NoStop}%
\bibitem [{\citenamefont {Golubev}(2017)}]{Golubev2017}%
  \BibitemOpen
  \bibfield  {author} {\bibinfo {author} {\bibfnamefont {A.}~\bibnamefont
  {Golubev}},\ }\bibfield  {title} {\bibinfo {title} {Exponentially modified
  peak functions in biomedical sciences and related disciplines},\ }\href
  {https://doi.org/10.1155/2017/7925106} {\bibfield  {journal} {\bibinfo
  {journal} {Comput. Math. Methods Med.}\ }\textbf {\bibinfo {volume} {2017}},\
  \bibinfo {pages} {7925106} (\bibinfo {year} {2017})}\BibitemShut {NoStop}%
\bibitem [{\citenamefont {Matzke}\ and\ \citenamefont
  {Wagenmakers}(2009)}]{Matzke2009}%
  \BibitemOpen
  \bibfield  {author} {\bibinfo {author} {\bibfnamefont {D.}~\bibnamefont
  {Matzke}}\ and\ \bibinfo {author} {\bibfnamefont {E.-J.}\ \bibnamefont
  {Wagenmakers}},\ }\bibfield  {title} {\bibinfo {title} {Psychological
  interpretation of the ex-gaussian and shifted wald parameters: A diffusion
  model analysis},\ }\href {https://doi.org/10.3758/pbr.16.5.798} {\bibfield
  {journal} {\bibinfo  {journal} {Psychon. Bull. Rev.}\ }\textbf {\bibinfo
  {volume} {16}},\ \bibinfo {pages} {798} (\bibinfo {year} {2009})}\BibitemShut
  {NoStop}%
\bibitem [{\citenamefont {Pan}\ \emph {et~al.}(2020)\citenamefont {Pan},
  \citenamefont {Fu}, \citenamefont {Spitha}, \citenamefont {Zhao},
  \citenamefont {Roy}, \citenamefont {Morrow}, \citenamefont {Kohler},
  \citenamefont {Wright},\ and\ \citenamefont {Jin}}]{Pan_Jin_2020}%
  \BibitemOpen
  \bibfield  {author} {\bibinfo {author} {\bibfnamefont {D.}~\bibnamefont
  {Pan}}, \bibinfo {author} {\bibfnamefont {Y.}~\bibnamefont {Fu}}, \bibinfo
  {author} {\bibfnamefont {N.}~\bibnamefont {Spitha}}, \bibinfo {author}
  {\bibfnamefont {Y.}~\bibnamefont {Zhao}}, \bibinfo {author} {\bibfnamefont
  {C.~R.}\ \bibnamefont {Roy}}, \bibinfo {author} {\bibfnamefont {D.~J.}\
  \bibnamefont {Morrow}}, \bibinfo {author} {\bibfnamefont {D.~D.}\
  \bibnamefont {Kohler}}, \bibinfo {author} {\bibfnamefont {J.~C.}\
  \bibnamefont {Wright}},\ and\ \bibinfo {author} {\bibfnamefont
  {S.}~\bibnamefont {Jin}},\ }\bibfield  {title} {\bibinfo {title}
  {Deterministic fabrication of arbitrary vertical heterostructures of
  two-dimensional ruddlesden{\textendash}popper halide perovskites},\ }\href
  {https://doi.org/10.1038/s41565-020-00802-2} {\bibfield  {journal} {\bibinfo
  {journal} {Nat. Nanotechnol.}\ }\textbf {\bibinfo {volume} {16}},\ \bibinfo
  {pages} {159} (\bibinfo {year} {2020})}\BibitemShut {NoStop}%
\bibitem [{\citenamefont {Elbaz}\ \emph {et~al.}(2017)\citenamefont {Elbaz},
  \citenamefont {Straus}, \citenamefont {Semonin}, \citenamefont {Hull},
  \citenamefont {Paley}, \citenamefont {Kim}, \citenamefont {Owen},
  \citenamefont {Kagan},\ and\ \citenamefont {Roy}}]{Elbaz2017}%
  \BibitemOpen
  \bibfield  {author} {\bibinfo {author} {\bibfnamefont {G.~A.}\ \bibnamefont
  {Elbaz}}, \bibinfo {author} {\bibfnamefont {D.~B.}\ \bibnamefont {Straus}},
  \bibinfo {author} {\bibfnamefont {O.~E.}\ \bibnamefont {Semonin}}, \bibinfo
  {author} {\bibfnamefont {T.~D.}\ \bibnamefont {Hull}}, \bibinfo {author}
  {\bibfnamefont {D.~W.}\ \bibnamefont {Paley}}, \bibinfo {author}
  {\bibfnamefont {P.}~\bibnamefont {Kim}}, \bibinfo {author} {\bibfnamefont
  {J.~S.}\ \bibnamefont {Owen}}, \bibinfo {author} {\bibfnamefont {C.~R.}\
  \bibnamefont {Kagan}},\ and\ \bibinfo {author} {\bibfnamefont
  {X.}~\bibnamefont {Roy}},\ }\bibfield  {title} {\bibinfo {title} {Unbalanced
  hole and electron diffusion in lead bromide perovskites},\ }\href
  {https://doi.org/10.1021/acs.nanolett.6b05022} {\bibfield  {journal}
  {\bibinfo  {journal} {Nano Lett.}\ }\textbf {\bibinfo {volume} {17}},\
  \bibinfo {pages} {1727} (\bibinfo {year} {2017})}\BibitemShut {NoStop}%
\bibitem [{\citenamefont {Lockwood}\ and\ \citenamefont
  {Wasilewski}(2004)}]{Lockwood2004}%
  \BibitemOpen
  \bibfield  {author} {\bibinfo {author} {\bibfnamefont {D.~J.}\ \bibnamefont
  {Lockwood}}\ and\ \bibinfo {author} {\bibfnamefont {Z.~R.}\ \bibnamefont
  {Wasilewski}},\ }\bibfield  {title} {\bibinfo {title} {Optical phonons in
  {Al}$_x${Ga}$_{1-x}${As}: Raman spectroscopy},\ }\href
  {https://doi.org/10.1103/physrevb.70.155202} {\bibfield  {journal} {\bibinfo
  {journal} {Phys. Rev. B}\ }\textbf {\bibinfo {volume} {70}},\ \bibinfo
  {pages} {155202} (\bibinfo {year} {2004})}\BibitemShut {NoStop}%
\bibitem [{\citenamefont {Ardekani}\ \emph {et~al.}(2019)\citenamefont
  {Ardekani}, \citenamefont {Younts}, \citenamefont {Yu}, \citenamefont {Cao},\
  and\ \citenamefont {Gundogdu}}]{Ardekani2019}%
  \BibitemOpen
  \bibfield  {author} {\bibinfo {author} {\bibfnamefont {H.}~\bibnamefont
  {Ardekani}}, \bibinfo {author} {\bibfnamefont {R.}~\bibnamefont {Younts}},
  \bibinfo {author} {\bibfnamefont {Y.}~\bibnamefont {Yu}}, \bibinfo {author}
  {\bibfnamefont {L.}~\bibnamefont {Cao}},\ and\ \bibinfo {author}
  {\bibfnamefont {K.}~\bibnamefont {Gundogdu}},\ }\bibfield  {title} {\bibinfo
  {title} {Reversible photoluminescence tuning by defect passivation via laser
  irradiation on aged monolayer {MoS}$_2$},\ }\href
  {https://doi.org/10.1021/acsami.9b10688} {\bibfield  {journal} {\bibinfo
  {journal} {{ACS} Appl. Mater. Interfaces}\ }\textbf {\bibinfo {volume}
  {11}},\ \bibinfo {pages} {38240} (\bibinfo {year} {2019})}\BibitemShut
  {NoStop}%
\bibitem [{\citenamefont {Yoo}\ \emph {et~al.}(2015)\citenamefont {Yoo},
  \citenamefont {Kang}, \citenamefont {Murai},\ and\ \citenamefont
  {Yoshimoto}}]{Yoo2015}%
  \BibitemOpen
  \bibfield  {author} {\bibinfo {author} {\bibfnamefont {W.~S.}\ \bibnamefont
  {Yoo}}, \bibinfo {author} {\bibfnamefont {K.}~\bibnamefont {Kang}}, \bibinfo
  {author} {\bibfnamefont {G.}~\bibnamefont {Murai}},\ and\ \bibinfo {author}
  {\bibfnamefont {M.}~\bibnamefont {Yoshimoto}},\ }\bibfield  {title} {\bibinfo
  {title} {Temperature dependence of photoluminescence spectra from crystalline
  silicon},\ }\href {https://doi.org/10.1149/2.0251512jss} {\bibfield
  {journal} {\bibinfo  {journal} {{ECS} J. Solid State Sci. Technol.}\ }\textbf
  {\bibinfo {volume} {4}},\ \bibinfo {pages} {P456} (\bibinfo {year}
  {2015})}\BibitemShut {NoStop}%
\bibitem [{\citenamefont {Zaghloul}\ and\ \citenamefont
  {Ali}(2011)}]{Zaghloul_Ali_2011}%
  \BibitemOpen
  \bibfield  {author} {\bibinfo {author} {\bibfnamefont {M.~R.}\ \bibnamefont
  {Zaghloul}}\ and\ \bibinfo {author} {\bibfnamefont {A.~N.}\ \bibnamefont
  {Ali}},\ }\bibfield  {title} {\bibinfo {title} {Algorithm 916: Computing the
  faddeyeva and voigt functions},\ }\href
  {https://doi.org/10.1145/2049673.2049679} {\bibfield  {journal} {\bibinfo
  {journal} {ACM Trans. Math. Softw.}\ }\textbf {\bibinfo {volume} {38}},\
  \bibinfo {pages} {1} (\bibinfo {year} {2011})}\BibitemShut {NoStop}%
\bibitem [{\citenamefont {Abrarov}\ and\ \citenamefont
  {Quine}(2018)}]{Abrarov_Quine_2018}%
  \BibitemOpen
  \bibfield  {author} {\bibinfo {author} {\bibfnamefont {S.~M.}\ \bibnamefont
  {Abrarov}}\ and\ \bibinfo {author} {\bibfnamefont {B.~M.}\ \bibnamefont
  {Quine}},\ }\bibfield  {title} {\bibinfo {title} {A rational approximation of
  the dawson's integral for efficient computation of the complex error
  function},\ }\href {https://doi.org/10.1016/j.amc.2017.10.032} {\bibfield
  {journal} {\bibinfo  {journal} {Appl. Math}\ }\textbf {\bibinfo {volume}
  {321}},\ \bibinfo {pages} {526} (\bibinfo {year} {2018})}\BibitemShut
  {NoStop}%
\bibitem [{\citenamefont {Grinvald}\ and\ \citenamefont
  {Steinberg}(1974)}]{Grinvald_Steinberg_1974}%
  \BibitemOpen
  \bibfield  {author} {\bibinfo {author} {\bibfnamefont {A.}~\bibnamefont
  {Grinvald}}\ and\ \bibinfo {author} {\bibfnamefont {I.~Z.}\ \bibnamefont
  {Steinberg}},\ }\bibfield  {title} {\bibinfo {title} {On the analysis of
  fluorescence decay kinetics by the method of least-squares},\ }\href
  {https://doi.org/10.1016/0003-2697(74)90312-1} {\bibfield  {journal}
  {\bibinfo  {journal} {Anal. Biochem.}\ }\textbf {\bibinfo {volume} {59}},\
  \bibinfo {pages} {583} (\bibinfo {year} {1974})}\BibitemShut {NoStop}%
\bibitem [{\citenamefont {{Van Den Zegel}}\ \emph {et~al.}(1986)\citenamefont
  {{Van Den Zegel}}, \citenamefont {Boens}, \citenamefont {Daems},\ and\
  \citenamefont {{De Schryver}}}]{VanDenZegel_DeSchryver_1986}%
  \BibitemOpen
  \bibfield  {author} {\bibinfo {author} {\bibfnamefont {M.}~\bibnamefont {{Van
  Den Zegel}}}, \bibinfo {author} {\bibfnamefont {N.}~\bibnamefont {Boens}},
  \bibinfo {author} {\bibfnamefont {D.}~\bibnamefont {Daems}},\ and\ \bibinfo
  {author} {\bibfnamefont {F.}~\bibnamefont {{De Schryver}}},\ }\bibfield
  {title} {\bibinfo {title} {Possibilities and limitations of the
  time-correlated single photon counting technique: a comparative study of
  correction methods for the wavelength dependence of the instrument response
  function},\ }\href {https://doi.org/10.1016/0301-0104(86)85096-0} {\bibfield
  {journal} {\bibinfo  {journal} {Chem. Phys.}\ }\textbf {\bibinfo {volume}
  {101}},\ \bibinfo {pages} {311} (\bibinfo {year} {1986})}\BibitemShut
  {NoStop}%
\bibitem [{\citenamefont {Li}\ \emph {et~al.}(2020)\citenamefont {Li},
  \citenamefont {Natakorn}, \citenamefont {Chen}, \citenamefont {Safar},
  \citenamefont {Cunningham}, \citenamefont {Tian},\ and\ \citenamefont
  {Li}}]{Li_Li_2020}%
  \BibitemOpen
  \bibfield  {author} {\bibinfo {author} {\bibfnamefont {Y.}~\bibnamefont
  {Li}}, \bibinfo {author} {\bibfnamefont {S.}~\bibnamefont {Natakorn}},
  \bibinfo {author} {\bibfnamefont {Y.}~\bibnamefont {Chen}}, \bibinfo {author}
  {\bibfnamefont {M.}~\bibnamefont {Safar}}, \bibinfo {author} {\bibfnamefont
  {M.}~\bibnamefont {Cunningham}}, \bibinfo {author} {\bibfnamefont
  {J.}~\bibnamefont {Tian}},\ and\ \bibinfo {author} {\bibfnamefont {D.~D.-U.}\
  \bibnamefont {Li}},\ }\bibfield  {title} {\bibinfo {title} {Investigations on
  average fluorescence lifetimes for visualizing multi-exponential decays},\
  }\href {https://doi.org/10.3389/fphy.2020.576862} {\bibfield  {journal}
  {\bibinfo  {journal} {Front. Phys.}\ }\textbf {\bibinfo {volume} {8}},\
  \bibinfo {pages} {447} (\bibinfo {year} {2020})}\BibitemShut {NoStop}%
\bibitem [{\citenamefont {Virtanen}\ \emph {et~al.}(2020)\citenamefont
  {Virtanen}, \citenamefont {Gommers}, \citenamefont {Oliphant}, \citenamefont
  {Haberland}, \citenamefont {Reddy}, \citenamefont {Cournapeau}, \citenamefont
  {Burovski}, \citenamefont {Peterson}, \citenamefont {Weckesser},
  \citenamefont {Bright}, \citenamefont {van~der Walt}, \citenamefont {Brett},
  \citenamefont {Wilson}, \citenamefont {Millman}, \citenamefont {Mayorov},
  \citenamefont {Nelson}, \citenamefont {Jones}, \citenamefont {Kern},
  \citenamefont {Larson}, \citenamefont {Carey}, \citenamefont {Polat},
  \citenamefont {Feng}, \citenamefont {Moore}, \citenamefont {VanderPlas},
  \citenamefont {Laxalde}, \citenamefont {Perktold}, \citenamefont {Cimrman},
  \citenamefont {Henriksen}, \citenamefont {Quintero}, \citenamefont {Harris},
  \citenamefont {Archibald}, \citenamefont {Ribeiro}, \citenamefont
  {Pedregosa},\ and\ \citenamefont {van Mulbregt}}]{Virtanen_Mulbreght_2020}%
  \BibitemOpen
  \bibfield  {author} {\bibinfo {author} {\bibfnamefont {P.}~\bibnamefont
  {Virtanen}}, \bibinfo {author} {\bibfnamefont {R.}~\bibnamefont {Gommers}},
  \bibinfo {author} {\bibfnamefont {T.~E.}\ \bibnamefont {Oliphant}}, \bibinfo
  {author} {\bibfnamefont {M.}~\bibnamefont {Haberland}}, \bibinfo {author}
  {\bibfnamefont {T.}~\bibnamefont {Reddy}}, \bibinfo {author} {\bibfnamefont
  {D.}~\bibnamefont {Cournapeau}}, \bibinfo {author} {\bibfnamefont
  {E.}~\bibnamefont {Burovski}}, \bibinfo {author} {\bibfnamefont
  {P.}~\bibnamefont {Peterson}}, \bibinfo {author} {\bibfnamefont
  {W.}~\bibnamefont {Weckesser}}, \bibinfo {author} {\bibfnamefont
  {J.}~\bibnamefont {Bright}}, \bibinfo {author} {\bibfnamefont {S.~J.}\
  \bibnamefont {van~der Walt}}, \bibinfo {author} {\bibfnamefont
  {M.}~\bibnamefont {Brett}}, \bibinfo {author} {\bibfnamefont
  {J.}~\bibnamefont {Wilson}}, \bibinfo {author} {\bibfnamefont {K.~J.}\
  \bibnamefont {Millman}}, \bibinfo {author} {\bibfnamefont {N.}~\bibnamefont
  {Mayorov}}, \bibinfo {author} {\bibfnamefont {A.~R.~J.}\ \bibnamefont
  {Nelson}}, \bibinfo {author} {\bibfnamefont {E.}~\bibnamefont {Jones}},
  \bibinfo {author} {\bibfnamefont {R.}~\bibnamefont {Kern}}, \bibinfo {author}
  {\bibfnamefont {E.}~\bibnamefont {Larson}}, \bibinfo {author} {\bibfnamefont
  {C.~J.}\ \bibnamefont {Carey}}, \bibinfo {author} {\bibfnamefont
  {{\.I}.}~\bibnamefont {Polat}}, \bibinfo {author} {\bibfnamefont
  {Y.}~\bibnamefont {Feng}}, \bibinfo {author} {\bibfnamefont {E.~W.}\
  \bibnamefont {Moore}}, \bibinfo {author} {\bibfnamefont {J.}~\bibnamefont
  {VanderPlas}}, \bibinfo {author} {\bibfnamefont {D.}~\bibnamefont {Laxalde}},
  \bibinfo {author} {\bibfnamefont {J.}~\bibnamefont {Perktold}}, \bibinfo
  {author} {\bibfnamefont {R.}~\bibnamefont {Cimrman}}, \bibinfo {author}
  {\bibfnamefont {I.}~\bibnamefont {Henriksen}}, \bibinfo {author}
  {\bibfnamefont {E.~A.}\ \bibnamefont {Quintero}}, \bibinfo {author}
  {\bibfnamefont {C.~R.}\ \bibnamefont {Harris}}, \bibinfo {author}
  {\bibfnamefont {A.~M.}\ \bibnamefont {Archibald}}, \bibinfo {author}
  {\bibfnamefont {A.~H.}\ \bibnamefont {Ribeiro}}, \bibinfo {author}
  {\bibfnamefont {F.}~\bibnamefont {Pedregosa}},\ and\ \bibinfo {author}
  {\bibfnamefont {P.}~\bibnamefont {van Mulbregt}},\ }\bibfield  {title}
  {\bibinfo {title} {{SciPy} 1.0: fundamental algorithms for scientific
  computing in python},\ }\href {https://doi.org/10.1038/s41592-019-0686-2}
  {\bibfield  {journal} {\bibinfo  {journal} {Nat. Methods}\ ,\ \bibinfo
  {pages} {261}} (\bibinfo {year} {2020})}\BibitemShut {NoStop}%
\bibitem [{\citenamefont {Harris}\ \emph {et~al.}(2020)\citenamefont {Harris},
  \citenamefont {Millman}, \citenamefont {van~der Walt}, \citenamefont
  {Gommers}, \citenamefont {Virtanen}, \citenamefont {Cournapeau},
  \citenamefont {Wieser}, \citenamefont {Taylor}, \citenamefont {Berg},
  \citenamefont {Smith}, \citenamefont {Kern}, \citenamefont {Picus},
  \citenamefont {Hoyer}, \citenamefont {van Kerkwijk}, \citenamefont {Brett},
  \citenamefont {Haldane}, \citenamefont {del R{\'{i}}o}, \citenamefont
  {Wiebe}, \citenamefont {Peterson}, \citenamefont {G{\'{e}}rard-Marchant},
  \citenamefont {Sheppard}, \citenamefont {Reddy}, \citenamefont {Weckesser},
  \citenamefont {Abbasi}, \citenamefont {Gohlke},\ and\ \citenamefont
  {Oliphant}}]{harris2020array}%
  \BibitemOpen
  \bibfield  {author} {\bibinfo {author} {\bibfnamefont {C.~R.}\ \bibnamefont
  {Harris}}, \bibinfo {author} {\bibfnamefont {K.~J.}\ \bibnamefont {Millman}},
  \bibinfo {author} {\bibfnamefont {S.~J.}\ \bibnamefont {van~der Walt}},
  \bibinfo {author} {\bibfnamefont {R.}~\bibnamefont {Gommers}}, \bibinfo
  {author} {\bibfnamefont {P.}~\bibnamefont {Virtanen}}, \bibinfo {author}
  {\bibfnamefont {D.}~\bibnamefont {Cournapeau}}, \bibinfo {author}
  {\bibfnamefont {E.}~\bibnamefont {Wieser}}, \bibinfo {author} {\bibfnamefont
  {J.}~\bibnamefont {Taylor}}, \bibinfo {author} {\bibfnamefont
  {S.}~\bibnamefont {Berg}}, \bibinfo {author} {\bibfnamefont {N.~J.}\
  \bibnamefont {Smith}}, \bibinfo {author} {\bibfnamefont {R.}~\bibnamefont
  {Kern}}, \bibinfo {author} {\bibfnamefont {M.}~\bibnamefont {Picus}},
  \bibinfo {author} {\bibfnamefont {S.}~\bibnamefont {Hoyer}}, \bibinfo
  {author} {\bibfnamefont {M.~H.}\ \bibnamefont {van Kerkwijk}}, \bibinfo
  {author} {\bibfnamefont {M.}~\bibnamefont {Brett}}, \bibinfo {author}
  {\bibfnamefont {A.}~\bibnamefont {Haldane}}, \bibinfo {author} {\bibfnamefont
  {J.~F.}\ \bibnamefont {del R{\'{i}}o}}, \bibinfo {author} {\bibfnamefont
  {M.}~\bibnamefont {Wiebe}}, \bibinfo {author} {\bibfnamefont
  {P.}~\bibnamefont {Peterson}}, \bibinfo {author} {\bibfnamefont
  {P.}~\bibnamefont {G{\'{e}}rard-Marchant}}, \bibinfo {author} {\bibfnamefont
  {K.}~\bibnamefont {Sheppard}}, \bibinfo {author} {\bibfnamefont
  {T.}~\bibnamefont {Reddy}}, \bibinfo {author} {\bibfnamefont
  {W.}~\bibnamefont {Weckesser}}, \bibinfo {author} {\bibfnamefont
  {H.}~\bibnamefont {Abbasi}}, \bibinfo {author} {\bibfnamefont
  {C.}~\bibnamefont {Gohlke}},\ and\ \bibinfo {author} {\bibfnamefont {T.~E.}\
  \bibnamefont {Oliphant}},\ }\bibfield  {title} {\bibinfo {title} {Array
  programming with {NumPy}},\ }\href
  {https://doi.org/10.1038/s41586-020-2649-2} {\bibfield  {journal} {\bibinfo
  {journal} {Nature}\ }\textbf {\bibinfo {volume} {585}},\ \bibinfo {pages}
  {357} (\bibinfo {year} {2020})}\BibitemShut {NoStop}%
\bibitem [{\citenamefont {van~der Walt}\ \emph {et~al.}(2011)\citenamefont
  {van~der Walt}, \citenamefont {Colbert},\ and\ \citenamefont
  {Varoquaux}}]{vanderWalt_Varoquaux_2011}%
  \BibitemOpen
  \bibfield  {author} {\bibinfo {author} {\bibfnamefont {S.}~\bibnamefont
  {van~der Walt}}, \bibinfo {author} {\bibfnamefont {S.~C.}\ \bibnamefont
  {Colbert}},\ and\ \bibinfo {author} {\bibfnamefont {G.}~\bibnamefont
  {Varoquaux}},\ }\bibfield  {title} {\bibinfo {title} {The {NumPy} array: A
  structure for efficient numerical computation},\ }\href
  {https://doi.org/10.1109/mcse.2011.37} {\bibfield  {journal} {\bibinfo
  {journal} {Comput. Sci. Eng.}\ }\textbf {\bibinfo {volume} {13}},\ \bibinfo
  {pages} {22} (\bibinfo {year} {2011})}\BibitemShut {NoStop}%
\bibitem [{\citenamefont {van Rossum}\ \emph {et~al.}(01  )\citenamefont {van
  Rossum} \emph {et~al.}}]{vanRossum_2001}%
  \BibitemOpen
  \bibfield  {author} {\bibinfo {author} {\bibfnamefont {G.}~\bibnamefont {van
  Rossum}} \emph {et~al.},\ }\href {http://www.python.org/} {\bibinfo {title}
  {{Python}}} (\bibinfo {year} {2001--}),\ \bibinfo {note} {[Online; accessed
  2022-09-15]}\BibitemShut {NoStop}%
\bibitem [{\citenamefont {Spitha}\ \emph {et~al.}(2020)\citenamefont {Spitha},
  \citenamefont {Kohler}, \citenamefont {Hautzinger}, \citenamefont {Li},
  \citenamefont {Jin},\ and\ \citenamefont {Wright}}]{Spitha_Wright_2020}%
  \BibitemOpen
  \bibfield  {author} {\bibinfo {author} {\bibfnamefont {N.}~\bibnamefont
  {Spitha}}, \bibinfo {author} {\bibfnamefont {D.~D.}\ \bibnamefont {Kohler}},
  \bibinfo {author} {\bibfnamefont {M.~P.}\ \bibnamefont {Hautzinger}},
  \bibinfo {author} {\bibfnamefont {J.}~\bibnamefont {Li}}, \bibinfo {author}
  {\bibfnamefont {S.}~\bibnamefont {Jin}},\ and\ \bibinfo {author}
  {\bibfnamefont {J.~C.}\ \bibnamefont {Wright}},\ }\bibfield  {title}
  {\bibinfo {title} {Discerning between exciton and free-carrier behaviors in
  ruddlesden{\textendash}popper perovskite quantum wells through kinetic
  modeling of photoluminescence dynamics},\ }\href
  {https://doi.org/10.1021/acs.jpcc.0c06345} {\bibfield  {journal} {\bibinfo
  {journal} {J. Phys. Chem. C}\ }\textbf {\bibinfo {volume} {124}},\ \bibinfo
  {pages} {17430} (\bibinfo {year} {2020})}\BibitemShut {NoStop}%
\bibitem [{\citenamefont {Manger}\ \emph {et~al.}(2017)\citenamefont {Manger},
  \citenamefont {Rowley}, \citenamefont {Fu}, \citenamefont {Foote},
  \citenamefont {Rea}, \citenamefont {Wood}, \citenamefont {Jin}, \citenamefont
  {Wright},\ and\ \citenamefont {Goldsmith}}]{Manger_Goldsmith_2017}%
  \BibitemOpen
  \bibfield  {author} {\bibinfo {author} {\bibfnamefont {L.~H.}\ \bibnamefont
  {Manger}}, \bibinfo {author} {\bibfnamefont {M.~B.}\ \bibnamefont {Rowley}},
  \bibinfo {author} {\bibfnamefont {Y.}~\bibnamefont {Fu}}, \bibinfo {author}
  {\bibfnamefont {A.~K.}\ \bibnamefont {Foote}}, \bibinfo {author}
  {\bibfnamefont {M.~T.}\ \bibnamefont {Rea}}, \bibinfo {author} {\bibfnamefont
  {S.~L.}\ \bibnamefont {Wood}}, \bibinfo {author} {\bibfnamefont
  {S.}~\bibnamefont {Jin}}, \bibinfo {author} {\bibfnamefont {J.~C.}\
  \bibnamefont {Wright}},\ and\ \bibinfo {author} {\bibfnamefont {R.~H.}\
  \bibnamefont {Goldsmith}},\ }\bibfield  {title} {\bibinfo {title} {Global
  analysis of perovskite photophysics reveals importance of geminate
  pathways},\ }\href {https://doi.org/10.1021/acs.jpcc.6b11547} {\bibfield
  {journal} {\bibinfo  {journal} {J. Phys. Chem. C}\ }\textbf {\bibinfo
  {volume} {121}},\ \bibinfo {pages} {1062} (\bibinfo {year}
  {2017})}\BibitemShut {NoStop}%
\bibitem [{\citenamefont {Brown}\ \emph {et~al.}(1986)\citenamefont {Brown},
  \citenamefont {Ridley},\ and\ \citenamefont {Rarity}}]{Brown_Rarity_1986}%
  \BibitemOpen
  \bibfield  {author} {\bibinfo {author} {\bibfnamefont {R.~G.~W.}\
  \bibnamefont {Brown}}, \bibinfo {author} {\bibfnamefont {K.~D.}\ \bibnamefont
  {Ridley}},\ and\ \bibinfo {author} {\bibfnamefont {J.~G.}\ \bibnamefont
  {Rarity}},\ }\bibfield  {title} {\bibinfo {title} {Characterization of
  silicon avalanche photodiodes for photon correlation measurements 1: Passive
  quenching},\ }\href {https://doi.org/10.1364/ao.25.004122} {\bibfield
  {journal} {\bibinfo  {journal} {Appl. Opt.}\ }\textbf {\bibinfo {volume}
  {25}},\ \bibinfo {pages} {4122} (\bibinfo {year} {1986})}\BibitemShut
  {NoStop}%
\bibitem [{\citenamefont {Ziarkash}\ \emph {et~al.}(2018)\citenamefont
  {Ziarkash}, \citenamefont {Joshi}, \citenamefont {Stip{\v{c}}evi{\'{c}}},\
  and\ \citenamefont {Ursin}}]{Ziarkash_Ursin_2018}%
  \BibitemOpen
  \bibfield  {author} {\bibinfo {author} {\bibfnamefont {A.~W.}\ \bibnamefont
  {Ziarkash}}, \bibinfo {author} {\bibfnamefont {S.~K.}\ \bibnamefont {Joshi}},
  \bibinfo {author} {\bibfnamefont {M.}~\bibnamefont {Stip{\v{c}}evi{\'{c}}}},\
  and\ \bibinfo {author} {\bibfnamefont {R.}~\bibnamefont {Ursin}},\ }\bibfield
   {title} {\bibinfo {title} {Comparative study of afterpulsing behavior and
  models in single photon counting avalanche photo diode detectors},\ }\href
  {https://doi.org/10.1038/s41598-018-23398-z} {\bibfield  {journal} {\bibinfo
  {journal} {Sci. Rep.}\ }\textbf {\bibinfo {volume} {8}},\ \bibinfo {pages}
  {5076} (\bibinfo {year} {2018})}\BibitemShut {NoStop}%
\end{thebibliography}%

\end{document}